\documentclass[12pt]{article}
\usepackage{graphicx}
\usepackage{amsmath}
\usepackage{amssymb}
\usepackage{caption2}
\begin{document}
\bibliographystyle{prsty}
\begin{center}
{\large {\bf \sc{  Mass spectrum of the scalar hidden
charm and bottom tetraquark states }}} \\[2mm]
Zhi-Gang Wang \footnote{E-mail,wangzgyiti@yahoo.com.cn.  }     \\
 Department of Physics, North China Electric Power University,
Baoding 071003, P. R. China
\end{center}

\begin{abstract}
In this article, we study the mass spectrum of the scalar hidden
charm and bottom tetraquark states  with the QCD sum rules. The
numerical results are compared with the corresponding ones from a
relativistic quark model based on a quasipotential approach in QCD.
The relevant  values from the constituent diquark model  based on
the constituent diquark masses and the spin-spin
 interactions are also discussed.
\end{abstract}



\section{Introduction}

The Babar, Belle, CLEO, D0, CDF and FOCUS collaborations have
discovered (or confirmed) a large number of charmonium-like states
$X(3940)$, $X(3872)$, $Y(4260)$, $Y(4008)$, $Y(3940)$, $Y(4325)$,
$Y(4360)$, $Y(4660)$, etc,  and revitalized  the interest  in the
spectroscopy of the charmonium states
\cite{review1,review2,review3,review4}. For a concise review of the
experimental situation of the new charmonium-like states, one can
consult Ref.\cite{Olsen2009}. Many possible  assignments for those
states have been suggested, such as multiquark states (irrespective
of the molecule  type and the diquark-antidiquark type), hybrid
states, charmonium states modified by nearby thresholds, threshold
cusps, etc \cite{review1,review2,review3,review4}. The observed
decay channels  are $J/\psi \pi^+ \pi^-$  or $\psi'\pi^+\pi^-$, an
essential ingredient
 for understanding the structures of those
mesons is whether or not  the $\pi\pi$ comes from a resonance state.

 The $Z^+(4430)$  observed in the  decay mode $\psi^\prime\pi^+$ by the
 Belle collaboration is the most
interesting subject \cite{Belle-z4430}.  We can distinguish the
multiquark states
 from the hybrids or charmonia with the criterion of
non-zero charge. The $Z^+(4430)$ can't be a pure $c\bar{c}$ state
due to the positive charge,   and  may be a $c\bar{c}u\bar{d}$
tetraquark state. The Babar collaboration did not confirm this
resonance \cite{Babar0811}, i.e. they observed  no significant
evidence for a $Z(4430)$ signal for any of the processes
investigated, neither in the total $J/\psi\pi$ or $\psi'\pi$ mass
distribution nor in the corresponding distributions for the regions
of $K\pi$ mass for which observation of the $Z(4430)$ signal was
reported.

In 2008, the Belle collaboration  reported the first observation of
two resonance-like structures (thereafter we will denote them as
$Z(4050)$ and $Z(4250)$ respectively) in the $\pi^+\chi_{c1}$
invariant mass distribution near $4.1 \,\rm{GeV}$ in the exclusive
decays $\bar{B}^0\to K^- \pi^+ \chi_{c1}$  \cite{Belle-chipi}. Their
quark contents must be some special combinations of the $c\bar{c}
u\bar{d}$, just like the $Z^+(4430)$, they can't be the conventional
mesons.  They may be the tetraquark states \cite{
Wang0807,Wang08072} or the molecular states
\cite{Xliu0808,Lee09,SLee08,GDing09}. The $Z(4050)$ and $Z(4250)$
lie about $(0.5-0.6)\,\rm{GeV}$ above the $\pi^+\chi_{c1}$
threshold, the decay $ Z \to \pi^+\chi_{c1}$ can take place with the
"fall-apart" mechanism and it is OZI super-allowed, which can take
into account the large total width naturally.

The spins of the $Z(4050)$ and $Z(4250)$  are not determined yet,
they can be scalar or vector mesons.  In
Refs.\cite{Wang0807,Wang08072}, we assume  that the hidden charm
mesons $Z(4050)$ and $Z(4250)$ are vector (and scalar) tetraquark
states, and study their masses with the QCD sum rules. The numerical
results indicate that the mass of the vector hidden charm tetraquark
state is about $M_{Z}=(5.12\pm0.15)\,\rm{GeV}$ or
$M_{Z}=(5.16\pm0.16)\,\rm{GeV}$, while the mass of the scalar hidden
charm tetraquark state
 is about $M_{Z}=(4.36\pm0.18)\,\rm{GeV}$. The scalar hidden charm tetraquark states may have
smaller masses than the corresponding vector states.

The mass is a fundamental parameter in describing a hadron, whether
or not there exist those hidden charm tetraquark configurations is
of great importance itself, because it provides a new opportunity
for a deeper understanding of the low energy QCD.

In this article, we study the mass spectrum of the scalar hidden
charm and bottom tetraquark states using  the QCD sum rules
\cite{SVZ79,Reinders85}.  In the QCD sum rules, the operator product
expansion is used to expand the time-ordered currents into a series
of quark and gluon condensates which parameterize the long distance
properties of  the QCD vacuum. Based on the quark-hadron duality, we
can obtain copious information about the hadronic parameters at the
phenomenological side \cite{SVZ79,Reinders85}.

The $Z(4050)$ and $Z(4250)$ can be tentatively identified as the
scalar hidden charm ($c\bar{c}$) tetraquark states, while the scalar
hidden bottom ($b\bar{b}$) tetraquark states may be observed at the
LHCb, where the $b\bar{b}$ pairs will be copiously produced. The
hidden charm and bottom tetraquark states ($Z$) have the symbolic
quark structures:
\begin{align}
  Z^+ = Q\bar{Q} u  \bar{d}  ;~~~~
  Z^0 = \frac{1}{\sqrt{2}}Q\bar{Q}&( u  \bar{u}-d  \bar{d})  ;~~~~
  Z^- =Q\bar{Q}d\bar{u}    ; \nonumber\\
  Z_s^+ = Q\bar{Q}u  \bar{s} ;~~~~
  Z_s^- =Q\bar{Q}  s\bar{u}  ;&~~~~
  Z_s^0 = Q\bar{Q}d  \bar{s} ;~~~~
  \overline Z_s^0 = Q\bar{Q}s\bar{d} ; \nonumber \\
  Z_\varphi= \frac{1}{\sqrt{2}} Q\bar{Q} (u\bar{u}+d\bar{d});
  &~~~~Z_\phi =  Q\bar{Q} s\bar{ s} \, ,
\end{align}
where the $Q$ denotes  the heavy quarks $c$ and $b$.

The colored objects (diquarks) in a  confining potential can result
in a copious spectrum, there maybe  exist  a series of orbital
angular momentum excitations. In the heavy quark limit, the $c$ (and
$b$) quark can be taken  as a static well potential, which binds
the light quark $q$ to form a diquark in the color antitriplet
channel. We take the diquarks as the basic constituents   following
Jaffe and Wilczek \cite{Jaffe2003,Jaffe2004}.  The heavy tetraquark
system could be described  by a double-well potential with  two
light quarks $q'\bar{q}$ lying in the two wells respectively.

The diquarks have  five Dirac tensor structures, scalar $C\gamma_5$,
pseudoscalar $C$, vector $C\gamma_\mu \gamma_5$, axial vector
$C\gamma_\mu $  and  tensor $C\sigma_{\mu\nu}$. The structures
$C\gamma_\mu $ and $C\sigma_{\mu\nu}$ are symmetric, the structures
$C\gamma_5$, $C$ and $C\gamma_\mu \gamma_5$ are antisymmetric. The
attractive interactions of one-gluon exchange  favor  formation of
the diquarks in  color antitriplet $\overline{3}_{ c}$, flavor
antitriplet $\overline{3}_{ f}$ and spin singlet $1_s$
\cite{GI1,GI2}.   In this article, we assume the scalar hidden charm
and bottom  mesons $Z$ consist of  the $C\gamma_5-C \gamma_5$ type
diquark structures rather than the $C-C $ type diquark structures,
and construct the interpolating currents:
\begin{eqnarray}
J_{Z^+}(x)&=& \epsilon^{ijk}\epsilon^{imn}u_j^T(x) C\gamma_5 Q_k(x)
\bar{Q}_m(x) \gamma_5  C \bar{d}_n^T(x)\, , \nonumber\\
J_{Z^0}(x)&=& \frac{\epsilon^{ijk}\epsilon^{imn}}{\sqrt{2}}
\left[u_j^T(x) C\gamma_5 Q_k(x) \bar{Q}_m(x) \gamma_5  C
\bar{u}_n^T(x)-(u\rightarrow d)\right]\, , \nonumber\\
J_{Z^+_s}(x)&=& \epsilon^{ijk}\epsilon^{imn}u_j^T(x) C\gamma_5
Q_k(x)\bar{Q}_m(x) \gamma_5  C \bar{s}_n^T(x)\, , \nonumber\\
J_{Z^0_s}(x)&=& \epsilon^{ijk}\epsilon^{imn}d_j^T(x) C\gamma_5
Q_k(x)\bar{Q}_m(x) \gamma_5  C \bar{s}_n^T(x)\, , \nonumber\\
J_{Z_\varphi}(x)&=& \frac{\epsilon^{ijk}\epsilon^{imn}}{\sqrt{2}}
\left[u_j^T(x) C\gamma_5 Q_k(x) \bar{Q}_m(x) \gamma_5  C
\bar{u}_n^T(x) +(u\rightarrow d)\right]\, , \nonumber\\
 J_{Z_\phi}(x)&=& \epsilon^{ijk}\epsilon^{imn}s_j^T(x) C\gamma_5
Q_k(x)\bar{Q}_m(x) \gamma_5  C \bar{s}_n^T(x)\, ,
\end{eqnarray}
where the $i$, $j$, $k$, $\cdots$  are color indexes. In the isospin
limit, the interpolating currents result in three distinct
expressions for the correlation functions $\Pi(p)$ , which are
characterized by the number of the $s$ quark they contain.

The article is arranged as follows:  we derive the QCD sum rules for
  the scalar hidden charm and bottom tetraquark states $Z$  in section 2; in section 3, numerical
results and discussions; section 4 is reserved for conclusion.

\section{QCD sum rules for  the scalar tetraquark states $Z$ }
In the following, we write down  the two-point correlation functions
$\Pi(p)$  in the QCD sum rules,
\begin{eqnarray}
\Pi(p)&=&i\int d^4x e^{ip \cdot x} \langle
0|T\left\{J(x)J^{\dagger}(0)\right\}|0\rangle \, ,
\end{eqnarray}
where the  $J(x)$ denotes the interpolating currents  $J_{Z^+}(x)$,
$J_{Z^0}(x)$, $J_{Z^+_s}(x)$, etc.

We can insert  a complete set of intermediate hadronic states with
the same quantum numbers as the current operator $J(x)$ into the
correlation functions  $\Pi(p)$  to obtain the hadronic
representation \cite{SVZ79,Reinders85}. After isolating the ground
state contribution from the pole term of the $Z$, we get the
following result,
\begin{eqnarray}
\Pi(p)&=&\frac{\lambda_{Z}^2}{M_{Z}^2-p^2} +\cdots \, \, ,
\end{eqnarray}
where the pole residue (or coupling) $\lambda_Z$ is defined by
\begin{eqnarray}
\lambda_{Z}  &=& \langle 0|J(0)|Z(p)\rangle \, .
\end{eqnarray}

 After performing the standard procedure of the QCD sum rules, we obtain the following  six  sum rules:
\begin{eqnarray}
\lambda_{i}^2 e^{-\frac{M_i^2}{M^2}}= \int_{\Delta_i}^{s^0_i} ds
\rho_i(s)e^{-\frac{s}{M^2}} \, ,
\end{eqnarray}
\begin{eqnarray}
\rho_{q\bar{q}}(s)&=&\frac{1}{512 \pi^6}
\int_{\alpha_{min}}^{\alpha_{max}}d\alpha
\int_{\beta_{min}}^{1-\alpha} d\beta
\alpha\beta(1-\alpha-\beta)^3(s-\widetilde{m}^2_Q)^2(7s^2-6s\widetilde{m}^2_Q+\widetilde{m}^4_Q)
\nonumber \\
&&+\frac{ m_Q\langle \bar{q}q\rangle}{16 \pi^4}
\int_{\alpha_{min}}^{\alpha_{max}}d\alpha
\int_{\beta_{min}}^{1-\alpha} d\beta
(1-\alpha-\beta)(\alpha+\beta)(s-\widetilde{m}^2_Q) (\widetilde{m}^2_Q-2s) \nonumber\\
&& +\frac{ m_Q\langle \bar{q}g_s\sigma Gq\rangle}{64 \pi^4}
\int_{\alpha_{min}}^{\alpha_{max}}d\alpha
\int_{\beta_{min}}^{1-\alpha} d\beta (\alpha+\beta)
(3s-2\widetilde{m}^2_Q)  \nonumber\\
&&+\frac{m_Q^2\langle \bar{q}q\rangle^2}{12 \pi^2}
\int_{\alpha_{mix}}^{\alpha_{max}} d\alpha +\frac{ m_Q^2  \langle
\bar{q}g_s \sigma Gq\rangle^2 }{192 \pi^2 M^6}
\int_{\alpha_{mix}}^{\alpha_{max}} d\alpha
\widetilde{\widetilde{m}}^4_Q\delta(s-\widetilde{\widetilde{m}}^2_Q)\nonumber\\
&&-\frac{m_Q^2  \langle \bar{q}q\rangle\langle \bar{q}g_s \sigma
Gq\rangle }{24 \pi^2} \int_{\alpha_{mix}}^{\alpha_{max}} d\alpha
\left[1+\frac{s}{M^2}
\right]\delta(s-\widetilde{\widetilde{m}}^2_Q)\, ,
\end{eqnarray}

\begin{eqnarray}
\rho_{q\bar{s}}(s)&=&\frac{1}{512 \pi^6}
\int_{\alpha_{min}}^{\alpha_{max}}d\alpha
\int_{\beta_{min}}^{1-\alpha} d\beta
\alpha\beta(1-\alpha-\beta)^3(s-\widetilde{m}^2_Q)^2(7s^2-6s\widetilde{m}^2_Q+\widetilde{m}^4_Q)
\nonumber \\
&&+\frac{ m_sm_Q}{256 \pi^6}
\int_{\alpha_{min}}^{\alpha_{max}}d\alpha
\int_{\beta_{min}}^{1-\alpha} d\beta \beta
(1-\alpha-\beta)^2(s-\widetilde{m}^2_Q)^2(5s-2\widetilde{m}^2_Q)   \nonumber\\
&&+\frac{ m_s\langle \bar{s}s\rangle}{32 \pi^4}
\int_{\alpha_{min}}^{\alpha_{max}}d\alpha
\int_{\beta_{min}}^{1-\alpha} d\beta \alpha \beta
(1-\alpha-\beta)(10s^2-12s\widetilde{m}^2_Q+3\widetilde{m}^4_Q)   \nonumber\\
&&+\frac{ m_Q\langle \bar{q}q\rangle}{16 \pi^4}
\int_{\alpha_{min}}^{\alpha_{max}}d\alpha
\int_{\beta_{min}}^{1-\alpha} d\beta \alpha
(1-\alpha-\beta)(s-\widetilde{m}^2_Q) (\widetilde{m}^2_Q-2s) \nonumber\\
&&+\frac{ m_Q\langle \bar{s}s\rangle}{16 \pi^4}
\int_{\alpha_{min}}^{\alpha_{max}}d\alpha
\int_{\beta_{min}}^{1-\alpha} d\beta \beta
(1-\alpha-\beta) (s-\widetilde{m}^2_Q) (\widetilde{m}^2_Q-2s) \nonumber\\
&& +\frac{ m_Q\langle \bar{q}g_s\sigma Gq\rangle}{64 \pi^4}
\int_{\alpha_{min}}^{\alpha_{max}}d\alpha
\int_{\beta_{min}}^{1-\alpha} d\beta \alpha
(3s-2\widetilde{m}^2_Q)  \nonumber\\
&& +\frac{ m_Q\langle \bar{s}g_s\sigma Gs\rangle}{64 \pi^4}
\int_{\alpha_{min}}^{\alpha_{max}}d\alpha
\int_{\beta_{min}}^{1-\alpha} d\beta \beta
(3s-2\widetilde{m}^2_Q)  \nonumber\\
&&+\frac{ m_s\langle \bar{s}g_s\sigma Gs\rangle}{32 \pi^4}
\int_{\alpha_{min}}^{\alpha_{max}}d\alpha
\int_{\beta_{min}}^{1-\alpha} d\beta \alpha \beta
\left[\widetilde{m}^2_Q-2s-\frac{s^2}{6}\delta(s-\widetilde{m}^2_Q)\right]   \nonumber\\
&& +\frac{ m_sm_Q^2\langle \bar{q}q\rangle}{16 \pi^4}
\int_{\alpha_{min}}^{\alpha_{max}}d\alpha
\int_{\beta_{min}}^{1-\alpha} d\beta
(\widetilde{m}^2_Q-s)  \nonumber\\
&&+\frac{m_Q^2\langle \bar{q}q\rangle \langle \bar{s}s\rangle}{12
\pi^2} \int_{\alpha_{mix}}^{\alpha_{max}} d\alpha
+\frac{m_sm_Q^2\langle \bar{q}g_s\sigma Gq\rangle }{64 \pi^4}
\int_{\alpha_{mix}}^{\alpha_{max}} d\alpha \nonumber\\
&&-\frac{m_sm_Q\langle \bar{q}q\rangle \langle \bar{s}s\rangle}{24
\pi^2} \int_{\alpha_{mix}}^{\alpha_{max}} d\alpha \alpha
\left[2+s\delta(s-\widetilde{\widetilde{m}}^2_Q) \right]\nonumber\\
&&-\frac{m_Q^2  [\langle \bar{q}q\rangle\langle \bar{s}g_s \sigma
Gs\rangle+\langle \bar{s}s\rangle\langle \bar{q}g_s \sigma
Gq\rangle] }{48 \pi^2} \int_{\alpha_{mix}}^{\alpha_{max}} d\alpha
\left[1+\frac{s}{M^2} \right]\delta(s-\widetilde{\widetilde{m}}^2_Q)\nonumber\\
&&+\frac{m_s m_Q  [2\langle \bar{q}q\rangle\langle \bar{s}g_s \sigma
Gs\rangle+3\langle \bar{s}s\rangle\langle \bar{q}g_s \sigma
Gq\rangle] }{144 \pi^2} \int_{\alpha_{mix}}^{\alpha_{max}} d\alpha
\alpha\nonumber\\
&&\left[1+\frac{s}{M^2} +\frac{s^2}{2M^4}\right]\delta(s-\widetilde{\widetilde{m}}^2_Q)\nonumber\\
&&+\frac{ m_Q^2 \langle \bar{q}g_s \sigma Gq\rangle \langle
\bar{s}g_s \sigma Gs\rangle }{192 \pi^2 M^6}
\int_{\alpha_{mix}}^{\alpha_{max}} d\alpha
\widetilde{\widetilde{m}}^4_Q\delta(s-\widetilde{\widetilde{m}}^2_Q)\,
,
\end{eqnarray}

\begin{eqnarray}
\rho_{s\bar{s}}(s)&=&\frac{1}{512 \pi^6}
\int_{\alpha_{min}}^{\alpha_{max}}d\alpha
\int_{\beta_{min}}^{1-\alpha} d\beta
\alpha\beta(1-\alpha-\beta)^3(s-\widetilde{m}^2_Q)^2(7s^2-6s\widetilde{m}^2_Q+\widetilde{m}^4_Q)
\nonumber \\
&&+\frac{ m_sm_Q}{256 \pi^6}
\int_{\alpha_{min}}^{\alpha_{max}}d\alpha
\int_{\beta_{min}}^{1-\alpha} d\beta (\alpha+\beta)
(1-\alpha-\beta)^2(s-\widetilde{m}^2_Q)^2(5s-2\widetilde{m}^2_Q)   \nonumber\\
&&+\frac{ m_s\langle \bar{s}s\rangle}{16 \pi^4}
\int_{\alpha_{min}}^{\alpha_{max}}d\alpha
\int_{\beta_{min}}^{1-\alpha} d\beta \alpha \beta
(1-\alpha-\beta)(10s^2-12s\widetilde{m}^2_Q+3\widetilde{m}^4_Q)   \nonumber\\
&&+\frac{ m_Q\langle \bar{s}s\rangle}{16 \pi^4}
\int_{\alpha_{min}}^{\alpha_{max}}d\alpha
\int_{\beta_{min}}^{1-\alpha} d\beta (\alpha+\beta)
(1-\alpha-\beta) (s-\widetilde{m}^2_Q) (\widetilde{m}^2_Q-2s) \nonumber\\
&& +\frac{ m_Q\langle \bar{s}g_s\sigma Gs\rangle}{64 \pi^4}
\int_{\alpha_{min}}^{\alpha_{max}}d\alpha
\int_{\beta_{min}}^{1-\alpha} d\beta (\alpha+\beta)
(3s-2\widetilde{m}^2_Q)  \nonumber\\
&&+\frac{ m_s\langle \bar{s}g_s\sigma Gs\rangle}{16 \pi^4}
\int_{\alpha_{min}}^{\alpha_{max}}d\alpha
\int_{\beta_{min}}^{1-\alpha} d\beta \alpha \beta
\left[\widetilde{m}^2_Q-2s-\frac{s^2}{6}\delta(s-\widetilde{m}^2_Q)\right]   \nonumber\\
&& +\frac{ m_sm_Q^2\langle \bar{s}s\rangle}{8 \pi^4}
\int_{\alpha_{min}}^{\alpha_{max}}d\alpha
\int_{\beta_{min}}^{1-\alpha} d\beta
(\widetilde{m}^2_Q-s)  \nonumber\\
&&+\frac{m_Q^2  \langle \bar{s}s\rangle^2}{12 \pi^2}
\int_{\alpha_{mix}}^{\alpha_{max}} d\alpha +\frac{m_sm_Q^2\langle
\bar{s}g_s\sigma Gs\rangle }{32 \pi^4}
\int_{\alpha_{mix}}^{\alpha_{max}} d\alpha \nonumber\\
&&-\frac{m_sm_Q  \langle \bar{s}s\rangle^2}{12 \pi^2}
\int_{\alpha_{mix}}^{\alpha_{max}} d\alpha \alpha
\left[2+s\delta(s-\widetilde{\widetilde{m}}^2_Q) \right]\nonumber\\
&&-\frac{m_Q^2  \langle \bar{s}s\rangle\langle \bar{s}g_s \sigma
Gs\rangle }{24 \pi^2} \int_{\alpha_{mix}}^{\alpha_{max}} d\alpha
\left[1+\frac{s}{M^2} \right]\delta(s-\widetilde{\widetilde{m}}^2_Q)\nonumber\\
&&+\frac{5m_s m_Q  \langle \bar{s}s\rangle\langle \bar{s}g_s \sigma
Gs\rangle }{144 \pi^2} \int_{\alpha_{mix}}^{\alpha_{max}} d\alpha
\left[1+\frac{s}{M^2} +\frac{s^2}{2M^4}\right]\delta(s-\widetilde{\widetilde{m}}^2_Q)\nonumber\\
&&+\frac{ m_Q^2  \langle \bar{s}g_s \sigma Gs\rangle^2 }{192 \pi^2
M^6} \int_{\alpha_{mix}}^{\alpha_{max}} d\alpha
\widetilde{\widetilde{m}}^4_Q\delta(s-\widetilde{\widetilde{m}}^2_Q)\,
,
\end{eqnarray}
where the $i$ denote the $c\bar{c}q\bar{q}$,
   $c\bar{c}q\bar{s}$, $c\bar{c}s\bar{s}$, $b\bar{b}q\bar{q}$,
   $b\bar{b}q\bar{s}$ and $b\bar{b}s\bar{s}$ channels, respectively;
      the $s_i^0$ are the corresponding continuum threshold parameters and the $M^2$ is the Borel parameter;
      $\alpha_{max}=\frac{1+\sqrt{1-\frac{4m_Q^2}{s}}}{2}$,
$\alpha_{min}=\frac{1-\sqrt{1-\frac{4m_Q^2}{s}}}{2}$,
$\beta_{min}=\frac{\alpha m_Q^2}{\alpha s -m_Q^2}$,
$\widetilde{m}_Q^2=\frac{(\alpha+\beta)m_Q^2}{\alpha\beta}$,
$\widetilde{\widetilde{m}}_Q^2=\frac{m_Q^2}{\alpha(1-\alpha)}$.
 The thresholds $\Delta_i$ can be sorted into three sets,  we introduce the $q\bar{q}$,
$q\bar{s}$ and $s\bar{s}$ to denote the light quark constituents in
the scalar tetraquark states to simplify the notations,
$\Delta_{q\bar{q}}=4m_Q^2$, $\Delta_{q\bar{s}}=(2m_Q+m_s)^2$,
$\Delta_{s\bar{s}}=4(m_Q+m_s)^2$.

 We carry out the operator
product expansion to the vacuum condensates adding up to
dimension-10. In calculation, we
 take  assumption of vacuum saturation for  high
dimension vacuum condensates, they  are always
 factorized to lower condensates with vacuum saturation in the QCD sum rules,
  factorization works well in  large $N_c$ limit.
In this article, we take into account the contributions from the
quark condensates,  mixed condensates, and neglect the contributions
from the gluon condensate. The contributions  from the gluon
condensates  are suppressed by large denominators and would not play
any significant roles for the light tetraquark states
\cite{Wang1,Wang2}, the heavy tetraquark state \cite{Wang08072} and
the  heavy molecular state \cite{Wang0904}.  There are many terms
involving the gluon condensate for the heavy tetraquark states and
heavy molecular states in the operator product expansion (one can
consult Refs.\cite{Wang08072,Wang0904} for example), we neglect the
gluon condensates for simplicity.  Furthermore, we neglect the terms
proportional to the $m_u$ and $m_d$, their contributions are of
minor importance.

 Differentiate  the Eq.(6) with respect to  $\frac{1}{M^2}$, then eliminate the
 pole residues $\lambda_{i}$, we can obtain the six sum rules for
 the masses  of the $Z$,
 \begin{eqnarray}
 M_i^2= \frac{\int_{\Delta_i}^{s^0_i} ds
\frac{d}{d(-1/M^2)}\rho_i(s)e^{-\frac{s}{M^2}}
}{\int_{\Delta_i}^{s^0_i} ds \rho_i(s)e^{-\frac{s}{M^2}}}\, .
\end{eqnarray}

\section{Numerical results and discussions}
The input parameters are taken to be the standard values $\langle
\bar{q}q \rangle=-(0.24\pm 0.01 \,\rm{GeV})^3$, $\langle \bar{s}s
\rangle=(0.8\pm 0.2 )\langle \bar{q}q \rangle$, $\langle
\bar{q}g_s\sigma Gq \rangle=m_0^2\langle \bar{q}q \rangle$, $\langle
\bar{s}g_s\sigma Gs \rangle=m_0^2\langle \bar{s}s \rangle$,
$m_0^2=(0.8 \pm 0.2)\,\rm{GeV}^2$,  $m_s=(0.14\pm0.01)\,\rm{GeV}$,
$m_c=(1.35\pm0.10)\,\rm{GeV}$ and $m_b=(4.8\pm0.1)\,\rm{GeV}$ at the
energy scale about $\mu=1\, \rm{GeV}$
\cite{SVZ79,Reinders85,Ioffe2005}.

In the conventional QCD sum rules \cite{SVZ79,Reinders85}, there are
two criteria (pole dominance and convergence of the operator product
expansion) for choosing  the Borel parameter $M^2$ and threshold
parameter $s_0$. The light tetraquark states can not satisfy the two
criteria, although it is not an indication non-existence of the
light tetraquark states (For detailed discussions about this
subject, one can consult Refs.\cite{Wang08072,Wang0708}). We impose
the two criteria on the heavy tetraquark states to choose the Borel
parameter $M^2$ and threshold parameter $s_0$.

If the resonance-like structures $Z(4050)$ and $Z(4250)$ observed by
the Belle collaboration  in the $\pi^+\chi_{c1}$ invariant mass
distribution near $4.1 \,\rm{GeV}$ in the exclusive decays
$\bar{B}^0\to K^- \pi^+ \chi_{c1}$  are scalar tetraquark states
\cite{Belle-chipi}, the threshold parameter can be tentatively taken
as $s^0_{q\bar{q}}=(4.248+0.5)^2\, \rm{GeV}^2\approx 23 \,
\rm{GeV}^2$ to take into account all possible contributions from the
ground states,  where we choose the energy gap between the ground
states and the first radial excited states to be $0.5\,\rm{GeV}$.
Taking into account the $SU(3)$ symmetry of the light flavor quarks,
we expect the threshold parameters $s^0_{q\bar{s}}$ and
$s^0_{s\bar{s}}$ are slightly  larger than  the $s^0_{q\bar{q}}$.
Furthermore, we take into account the mass difference between the
$c$ and $b$ quarks, the threshold parameters in the   hidden bottom
channels are tentatively taken as $s^0_{q\bar{q}}=138\, \rm{GeV}^2$,
$s^0_{q\bar{s}}=140\, \rm{GeV}^2$ and $s^0_{s\bar{s}}=142\,
\rm{GeV}^2$.

Here we take it for granted that the energy gap between the ground
states and the first radial excited states is about $0.5\,\rm{GeV}$,
and use those  values as a guide to determine the threshold
parameters $s_0$ with the QCD sum rules.

The contributions from the high dimension vacuum condensates  in the
operator product expansion are shown in Figs.1-2, where (and
thereafter) we
 use the $\langle\bar{q}q\rangle$ to denote the quark condensates
$\langle\bar{q}q\rangle$, $\langle\bar{s}s\rangle$ and the
$\langle\bar{q}g_s \sigma Gq\rangle$ to denote the mixed condensates
$\langle\bar{q}g_s \sigma Gq\rangle$, $\langle\bar{s}g_s \sigma
Gs\rangle$. From the figures, we can see that the contributions from
the high dimension condensates change  quickly with variation of the
Borel parameter at the values $M^2\leq 2.6\,\rm{GeV}^2$ and $M^2\leq
7.0\,\rm{GeV}^2$ for the $c\bar{c}$ channels and $b\bar{b}$ channels
respectively, such an  unstable  behavior  can not lead to stable
sum rules, our numerical results confirm this conjecture.  At the
values $M^2\geq 2.6\,\rm{GeV}^2$ and $s_0\geq 23\,\rm{GeV}^2$, the
contributions from the  $\langle \bar{q}q\rangle^2+\langle
\bar{q}q\rangle \langle \bar{q}g_s \sigma Gq\rangle $ term are less
than (or equal) $10\%$ for the $c\bar{c}q\bar{q}$ channel, the
corresponding contributions are  smaller for the $c\bar{c}q\bar{s}$
and $c\bar{c}s\bar{s}$ channels; the contributions from the vacuum
condensate of the highest dimension $\langle\bar{q}g_s \sigma
Gq\rangle^2$ are less than (or equal) $2\%$ for all the $c\bar{c}$
channels, we expect the operator product expansion is convergent in
the $c\bar{c}$ channels. At the values $M^2\geq 7.0\,\rm{GeV}^2$ and
$s_0\geq 136\,\rm{GeV}^2$, the contributions from the  $\langle
\bar{q}q\rangle^2+\langle \bar{q}q\rangle \langle \bar{q}g_s \sigma
Gq\rangle $ term are less than  $10\%$ for the $b\bar{b}q\bar{q}$
channel, the corresponding contributions are  smaller for the
$b\bar{b}q\bar{s}$ and $b\bar{b}s\bar{s}$ channels; the
contributions from the vacuum condensate of the highest dimension
$\langle\bar{q}g_s \sigma Gq\rangle^2$ are less than (or equal)
$6\%$ for all the $b\bar{b}$ channels, we expect the operator
product expansion is convergent in the $b\bar{b}$ channels.

In Fig.3, we plot the contributions from different terms in the
operator product expansion.  From the figures, we can see that the
main contributions come from the perturbative term and the $\langle
\bar{q}q\rangle+ \langle \bar{q}g_s \sigma Gq\rangle $ term, the
operator product expansion is convergent; and the interpolating
currents contain more $s$ quarks have better convergent behavior.

 In this article, we take the uniform Borel parameter
$M^2_{min}$, i.e. $M^2_{min}\geq 2.6 \, \rm{GeV}^2$ and
$M^2_{min}\geq 7.0 \, \rm{GeV}^2$ for the $c\bar{c}$ channels and
$b\bar{b}$ channels, respectively.

\begin{figure}
 \centering
 \includegraphics[totalheight=5cm,width=6cm]{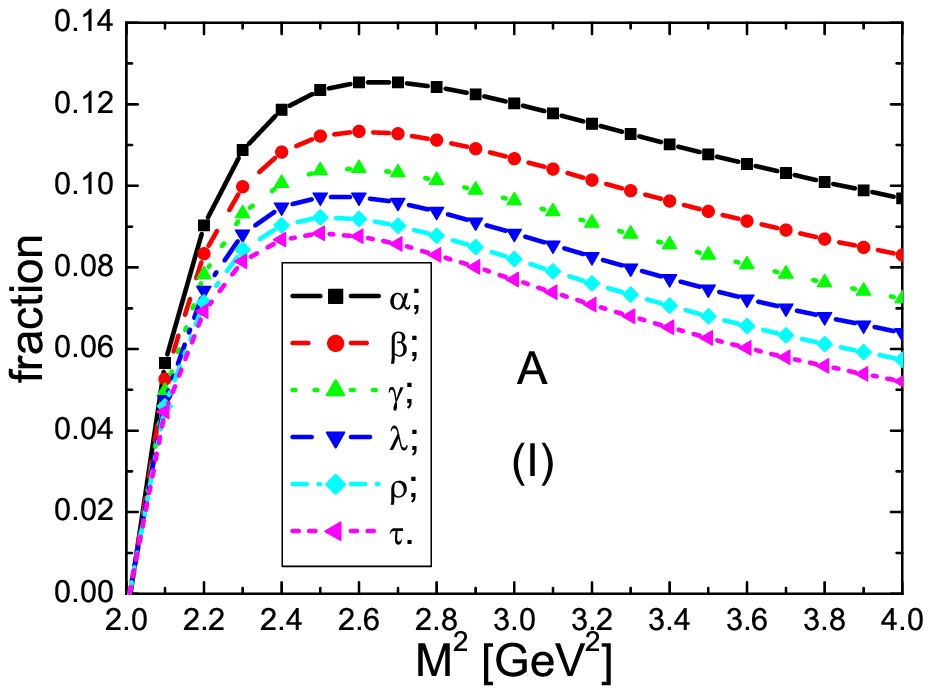}
 \includegraphics[totalheight=5cm,width=6cm]{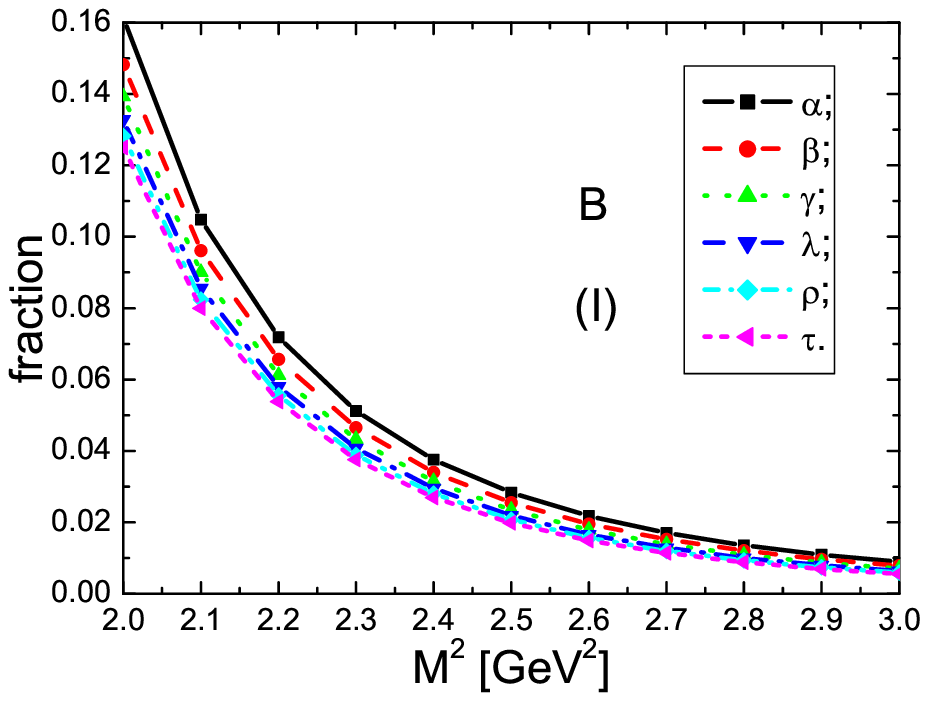}
 \includegraphics[totalheight=5cm,width=6cm]{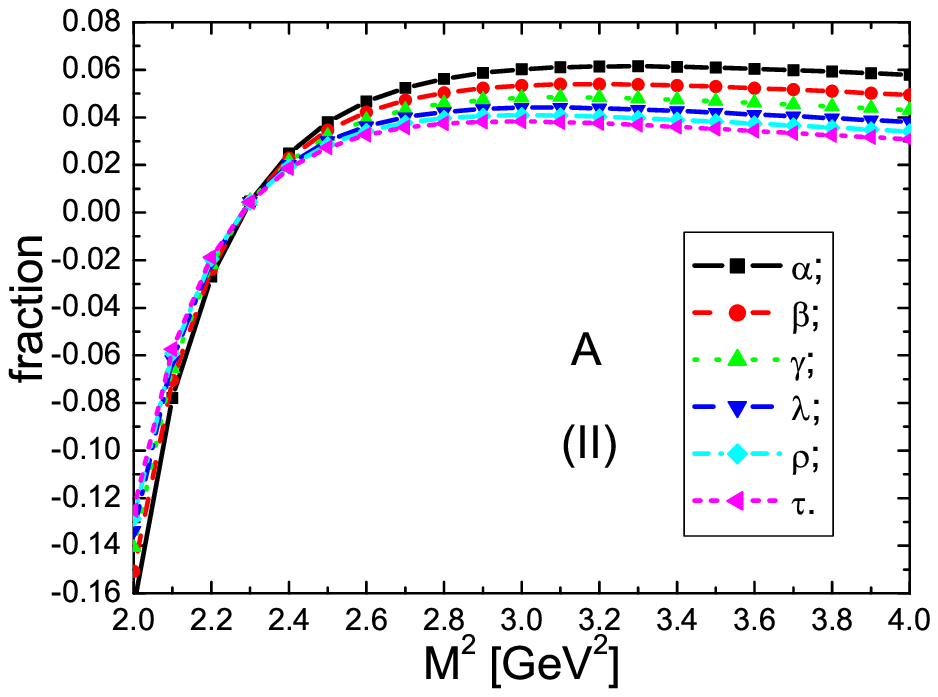}
 \includegraphics[totalheight=5cm,width=6cm]{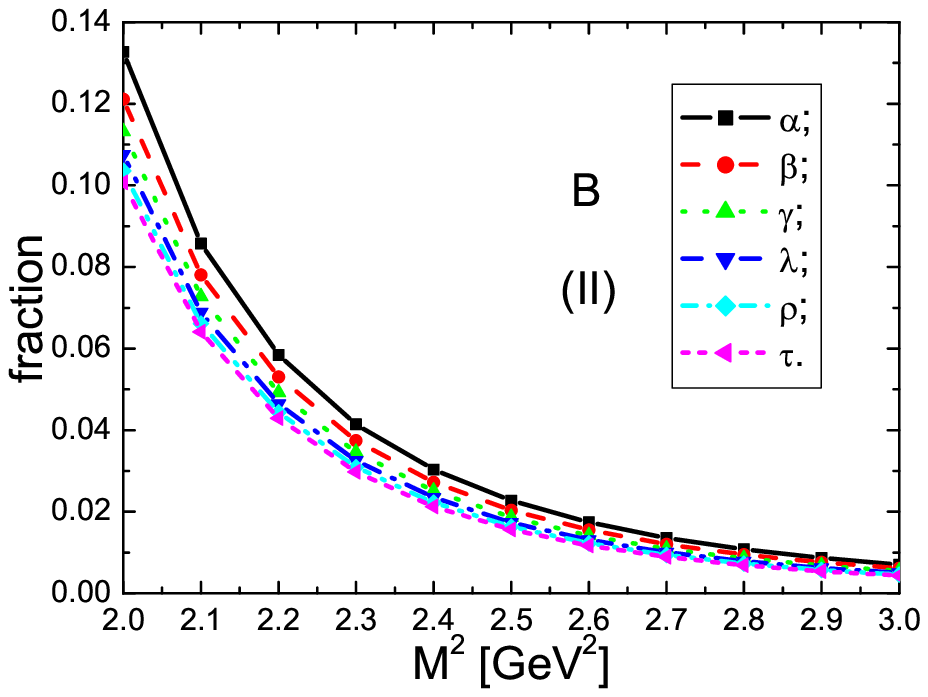}
 \includegraphics[totalheight=5cm,width=6cm]{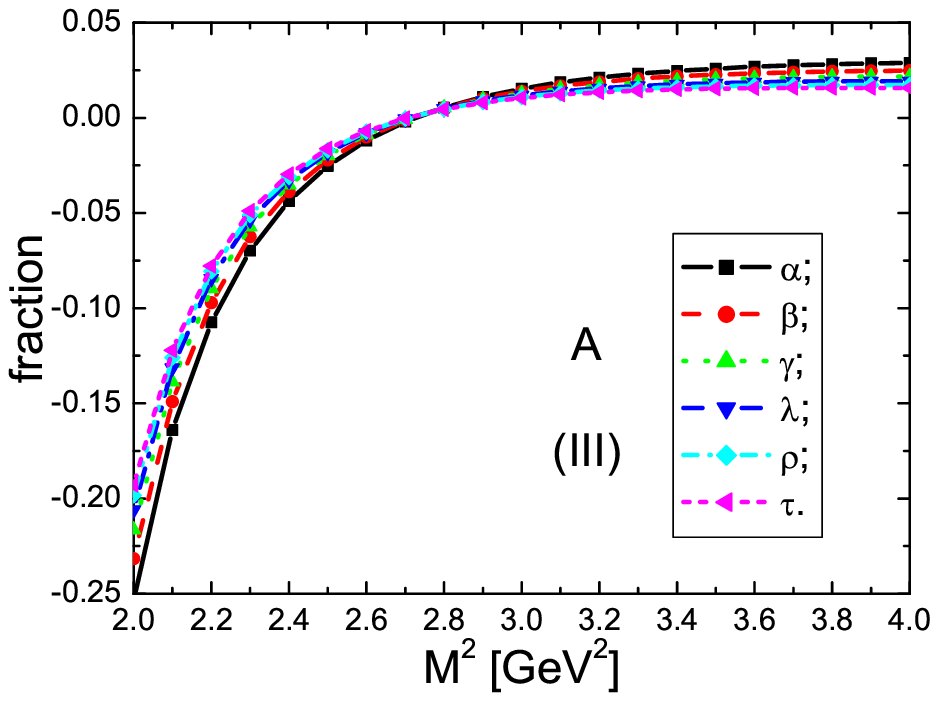}
 \includegraphics[totalheight=5cm,width=6cm]{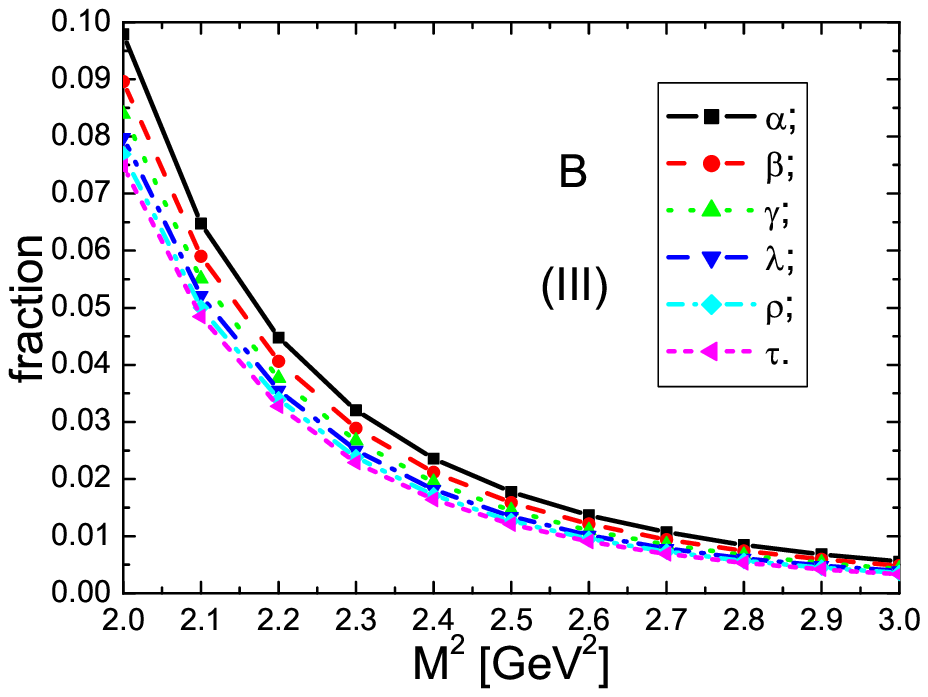}
   \caption{ The contributions from different terms with variation of the Borel
   parameter $M^2$  in the operator product expansion. The $A$ and
   $B$ denote the contributions from the
   $\langle \bar{q}q\rangle^2+\langle \bar{q}q\rangle \langle \bar{q}g_s \sigma Gq\rangle
   $ term and the  $ \langle \bar{q}g_s \sigma Gq\rangle^2
   $ term,  respectively. The
   (I), (II) and (III) denote the $c\bar{c}q\bar{q}$,
   $c\bar{c}q\bar{s}$ and $c\bar{c}s\bar{s}$ channels, respectively. The notations
   $\alpha$, $\beta$, $\gamma$, $\lambda$, $\rho$ and $\tau$  correspond to the threshold
   parameters $s_0=21\,\rm{GeV}^2$,
   $22\,\rm{GeV}^2$, $23\,\rm{GeV}^2$, $24\,\rm{GeV}^2$, $25\,\rm{GeV}^2$ and $26\,\rm{GeV}^2$, respectively.}
\end{figure}

\begin{figure}
 \centering
 \includegraphics[totalheight=5cm,width=6cm]{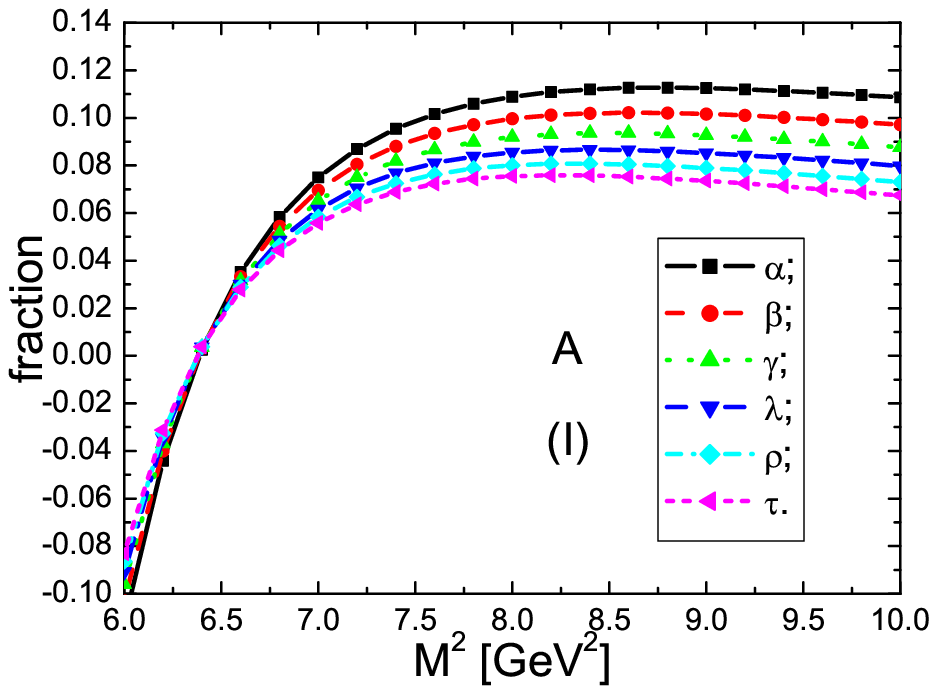}
 \includegraphics[totalheight=5cm,width=6cm]{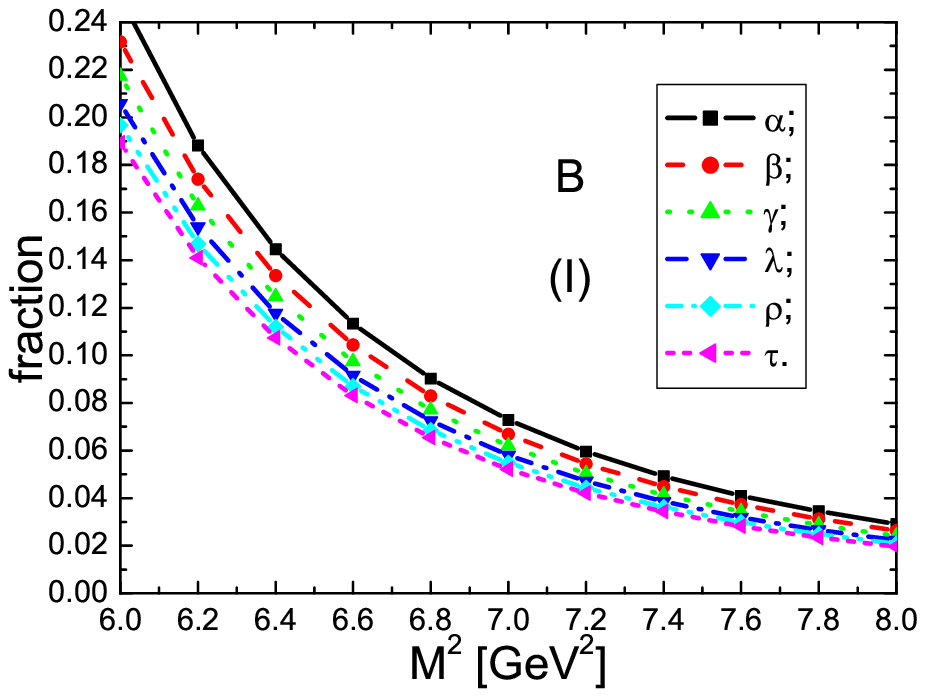}
 \includegraphics[totalheight=5cm,width=6cm]{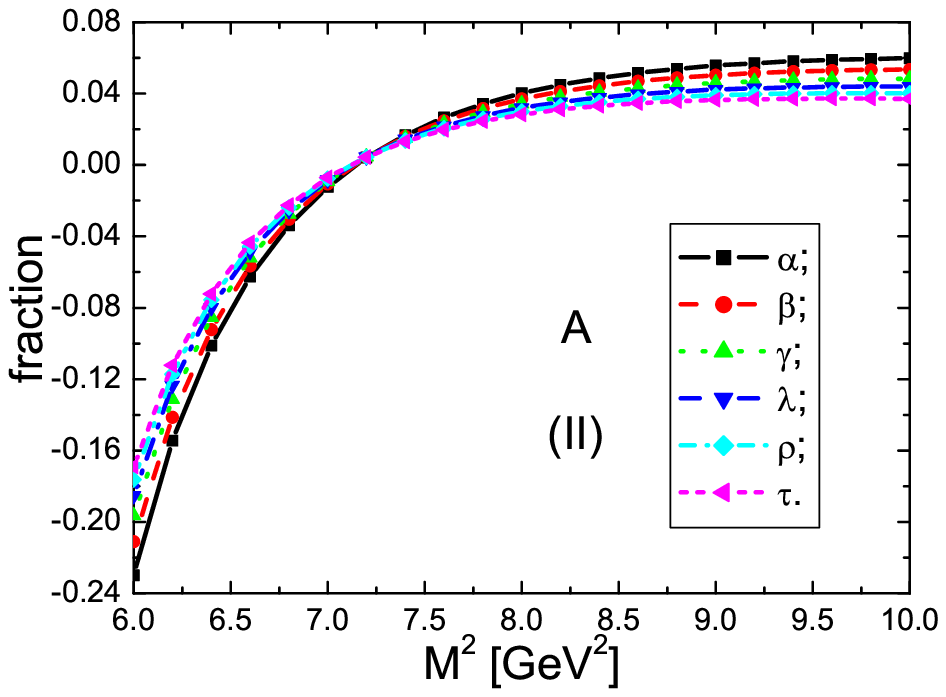}
 \includegraphics[totalheight=5cm,width=6cm]{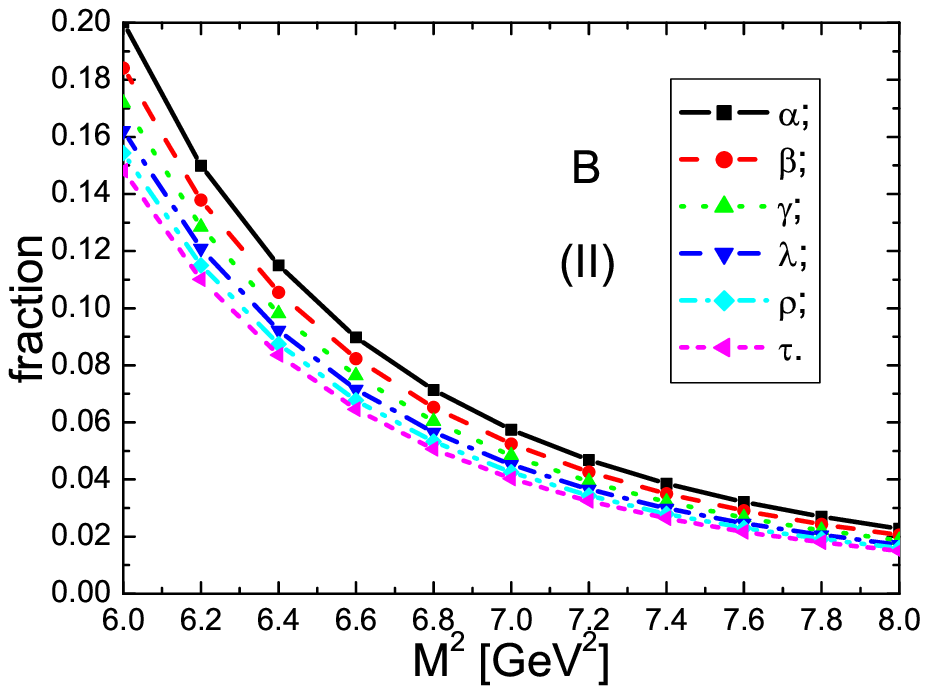}
 \includegraphics[totalheight=5cm,width=6cm]{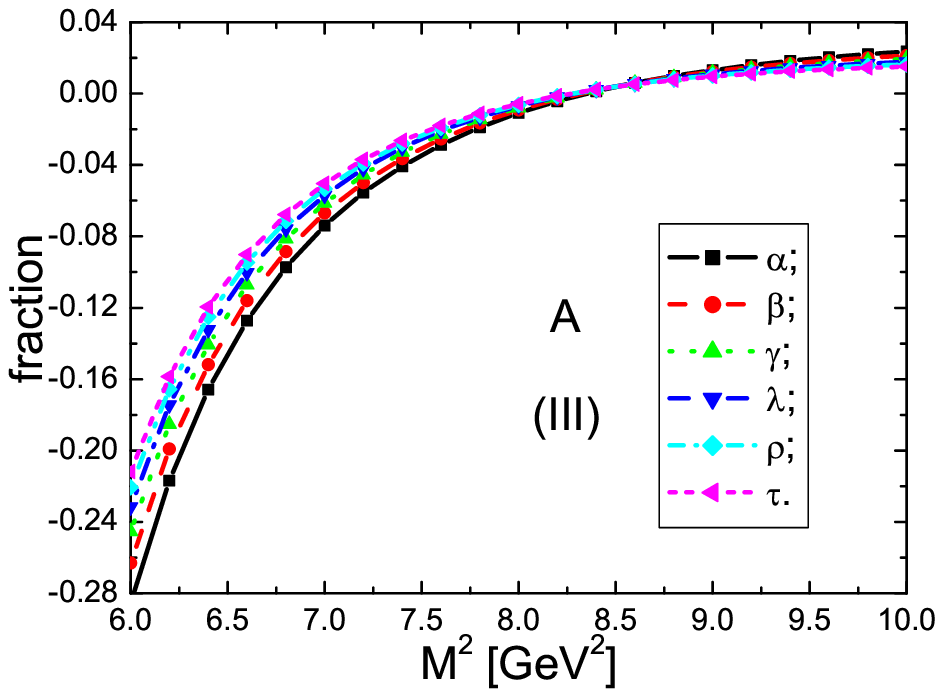}
 \includegraphics[totalheight=5cm,width=6cm]{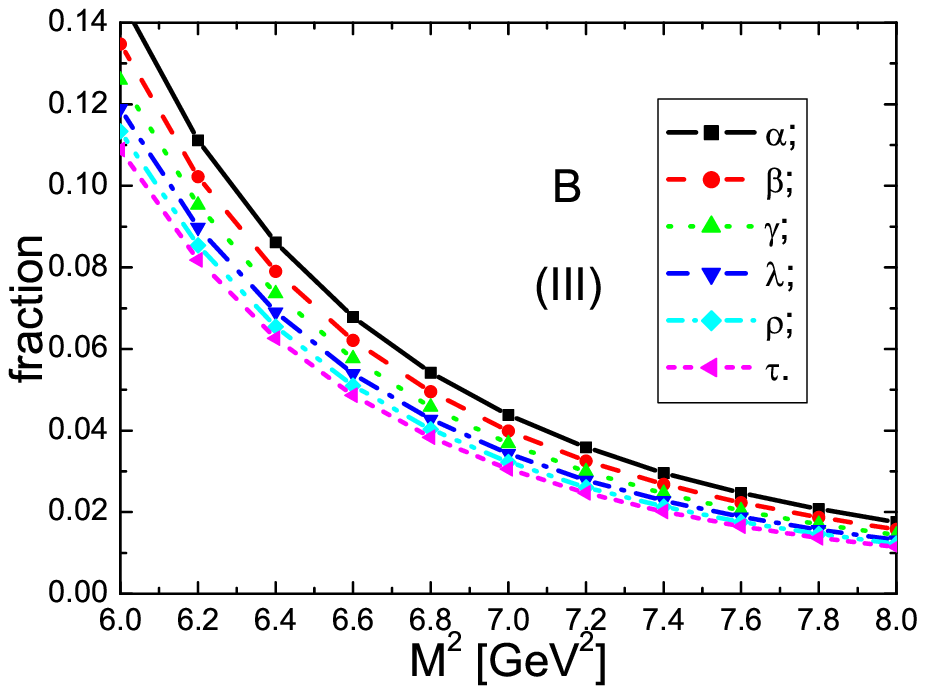}
   \caption{ The contributions from different terms with variation of the Borel
   parameter $M^2$  in the operator product expansion. The $A$ and
   $B$ denote the contributions from the
   $\langle \bar{q}q\rangle^2+\langle \bar{q}q\rangle \langle \bar{q}g_s \sigma Gq\rangle
   $ term and the  $ \langle \bar{q}g_s \sigma Gq\rangle^2
   $ term,  respectively. The
   (I), (II) and (III) denote the $b\bar{b}q\bar{q}$,
   $b\bar{b}q\bar{s}$ and $b\bar{b}s\bar{s}$ channels, respectively. The notations
   $\alpha$, $\beta$, $\gamma$, $\lambda$, $\rho$ and $\tau$  correspond to the threshold
   parameters $s_0=132\,\rm{GeV}^2$,
   $134\,\rm{GeV}^2$, $136\,\rm{GeV}^2$, $138\,\rm{GeV}^2$, $140\,\rm{GeV}^2$ and $142\,\rm{GeV}^2$, respectively. }
\end{figure}

\begin{figure}
 \centering
 \includegraphics[totalheight=5cm,width=6cm]{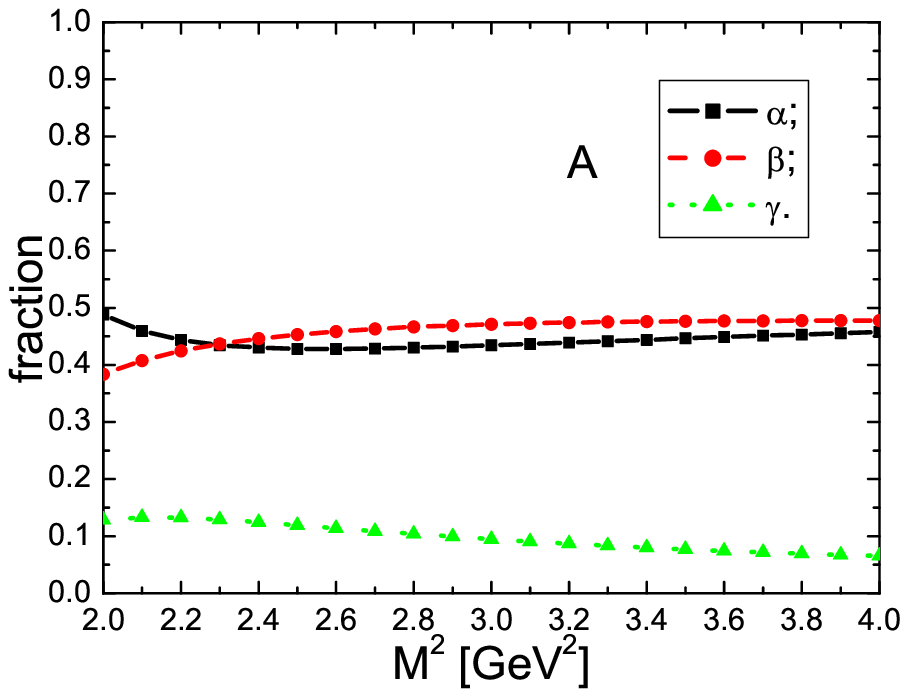}
 \includegraphics[totalheight=5cm,width=6cm]{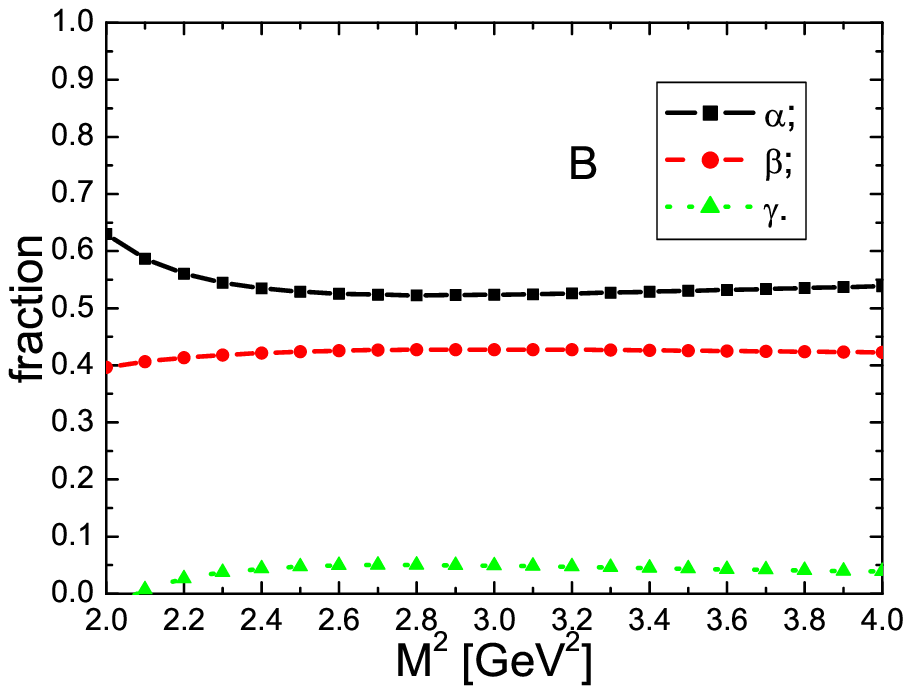}
 \includegraphics[totalheight=5cm,width=6cm]{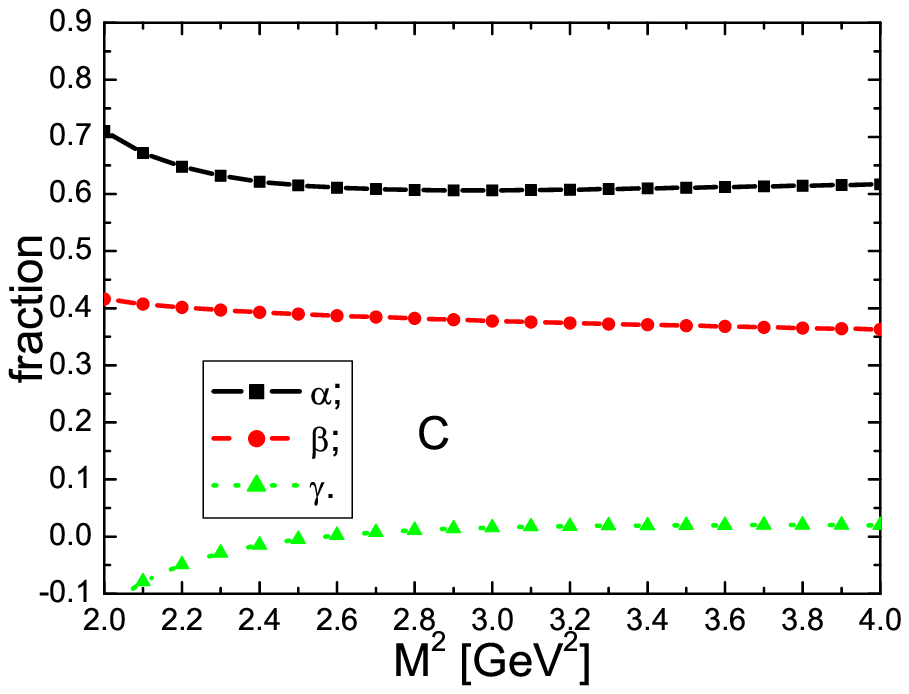}
 \includegraphics[totalheight=5cm,width=6cm]{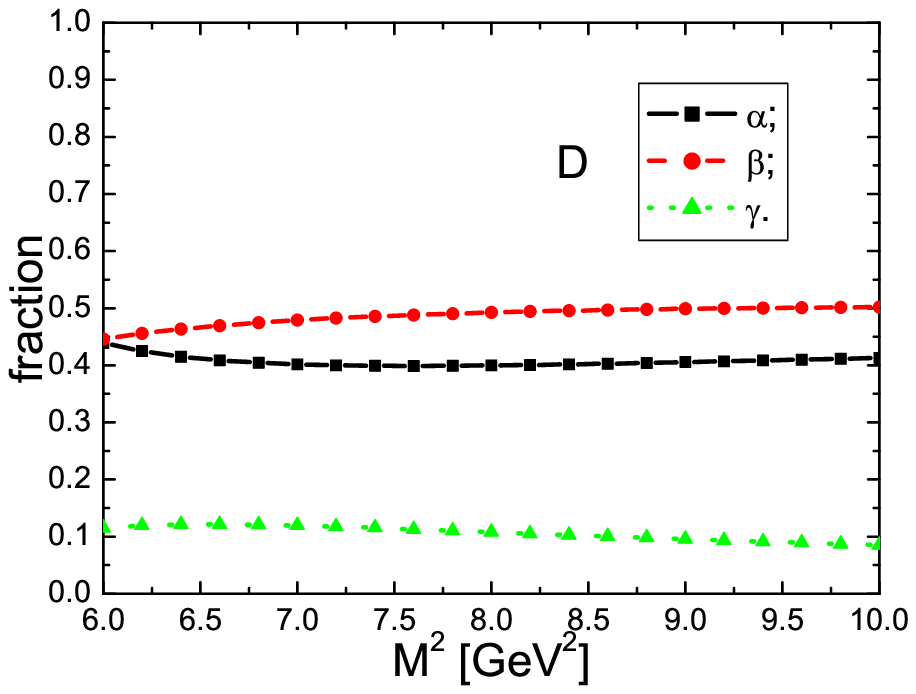}
 \includegraphics[totalheight=5cm,width=6cm]{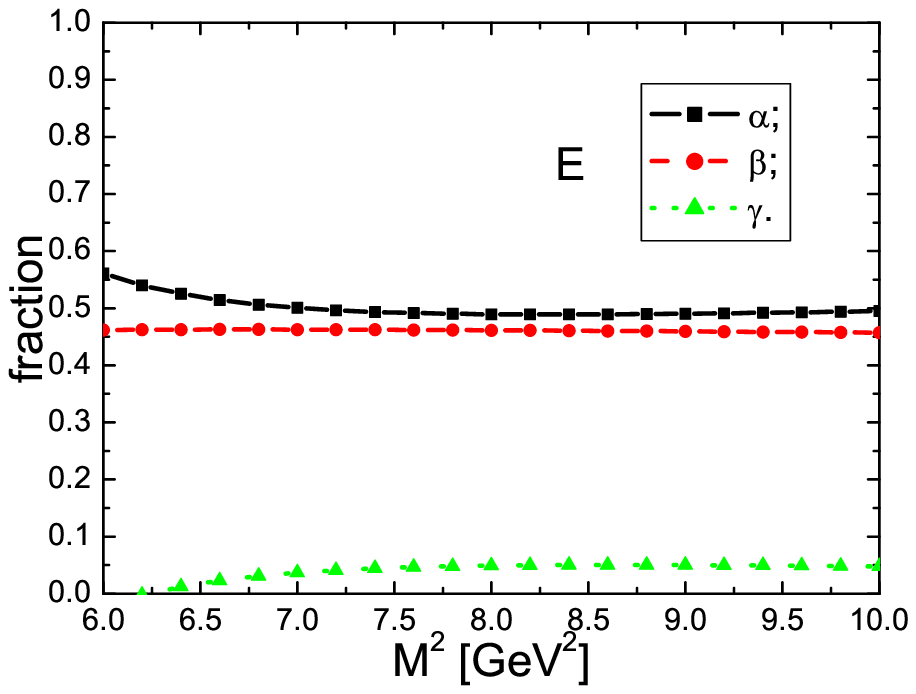}
 \includegraphics[totalheight=5cm,width=6cm]{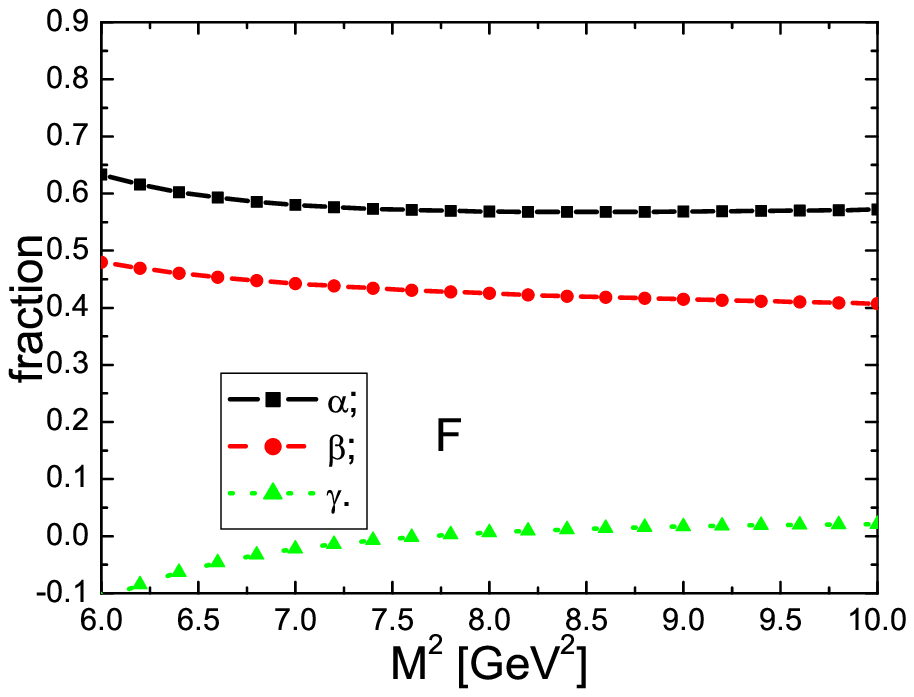}
   \caption{ The contributions from different terms with variation of the Borel
   parameter $M^2$  in the operator product expansion. The $A$, $B$, $C$,
   $D$, $E$ and $F$ denote the $c\bar{c}q\bar{q}$,
   $c\bar{c}q\bar{s}$, $c\bar{c}s\bar{s}$, $b\bar{b}q\bar{q}$,
   $b\bar{b}q\bar{s}$ and $b\bar{b}s\bar{s}$ channels, respectively.
   The $\alpha$, $\beta$ and $\gamma$
   correspond    to the perturbative term, the
    $\langle \bar{q}q\rangle+ \langle \bar{q}g_s \sigma Gq\rangle
   $  term and the
   $\langle \bar{q}q\rangle^2+\langle \bar{q}q\rangle \langle \bar{q}g_s \sigma
   Gq\rangle+\langle \bar{q}g_s \sigma Gq\rangle^2
   $ term, respectively. The threshold parameters are $s_0=24\,\rm{GeV}^2$ and $138\,\rm{GeV}^2$ for
   the $c\bar{c}$ channels and $b\bar{b}$ channels, respectively. }
\end{figure}

In Fig.4, we show the  contributions from the pole terms with
variation of the Borel parameter and the threshold parameter. The
pole contributions are larger than (or equal) $50\%$ at the value
$M^2 \leq 3.2 \, \rm{GeV}^2 $ and $s_0\geq
23\,\rm{GeV}^2,\,23\,\rm{GeV}^2,\,24\,\rm{GeV}^2$ for the
$c\bar{c}q\bar{q}$,
   $c\bar{c}q\bar{s}$, $c\bar{c}s\bar{s}$
channels respectively, and larger than (or equal) $50\%$ at the
value $M^2 \leq 8.0 \, \rm{GeV}^2 $ and $s_0\geq
136\,\rm{GeV}^2,\,138\,\rm{GeV}^2,\,138\,\rm{GeV}^2$ for  the
$b\bar{b}q\bar{q}$,
   $b\bar{b}q\bar{s}$ and $b\bar{b}s\bar{s}$ channels respectively. Again we
take the uniform Borel parameter $M^2_{max}$, i.e. $M^2_{max}\leq
3.2 \, \rm{GeV}^2$ and $M^2_{max}\leq 8.0 \, \rm{GeV}^2$ for the
$c\bar{c}$ channels and $b\bar{b}$ channels, respectively.

In this article, the threshold parameters are taken as
$s_0=(24\pm1)\,\rm{GeV}^2$, $(24\pm1)\,\rm{GeV}^2$,
$(25\pm1)\,\rm{GeV}^2$, $(138\pm2)\,\rm{GeV}^2$,
$(140\pm2)\,\rm{GeV}^2$ and $(140\pm2)\,\rm{GeV}^2$ for the
$c\bar{c}q\bar{q}$,
   $c\bar{c}q\bar{s}$, $c\bar{c}s\bar{s}$, $b\bar{b}q\bar{q}$,
   $b\bar{b}q\bar{s}$ and $b\bar{b}s\bar{s}$ channels, respectively;
   the Borel parameters are taken as $M^2=(2.6-3.2)\,\rm{GeV}^2$ and
   $(7.0-8.0)\,\rm{GeV}^2$ for the
$c\bar{c}$ channels and $b\bar{b}$ channels, respectively.
      In those regions, the two criteria of the QCD sum rules
are full satisfied  \cite{SVZ79,Reinders85}.

\begin{figure}
 \centering
 \includegraphics[totalheight=5cm,width=6cm]{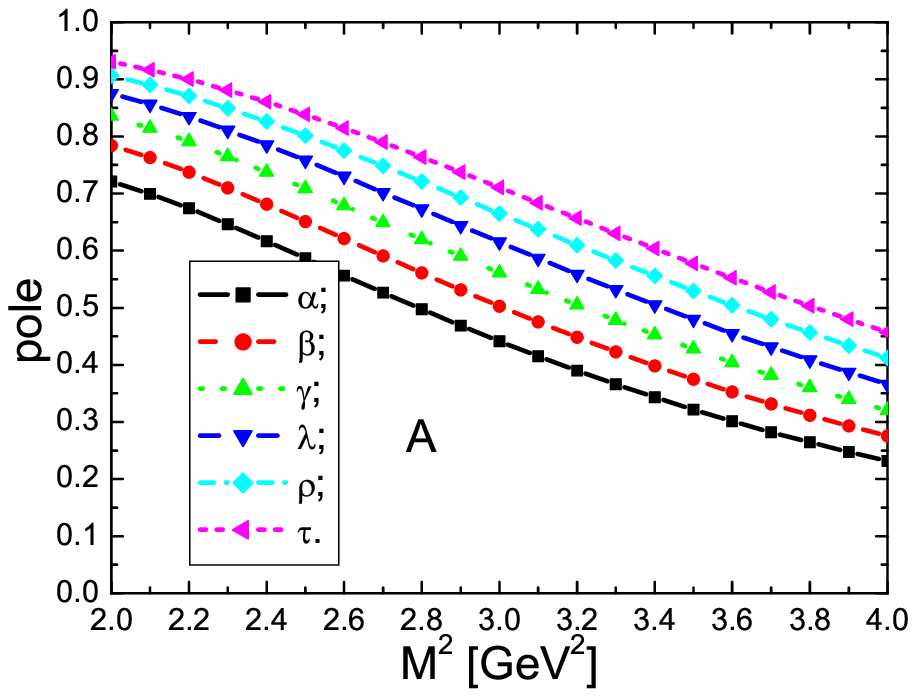}
 \includegraphics[totalheight=5cm,width=6cm]{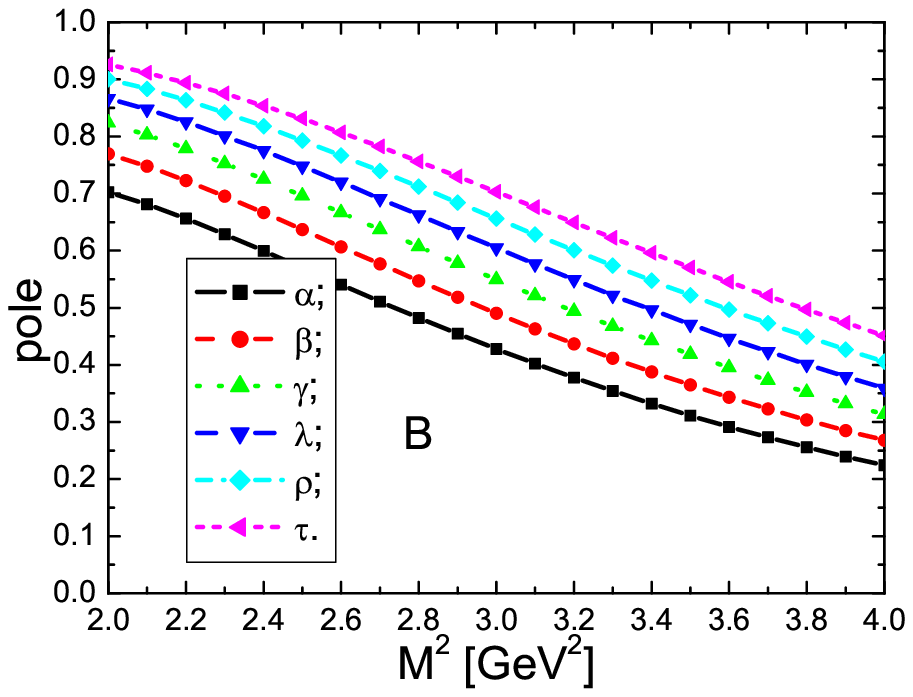}
 \includegraphics[totalheight=5cm,width=6cm]{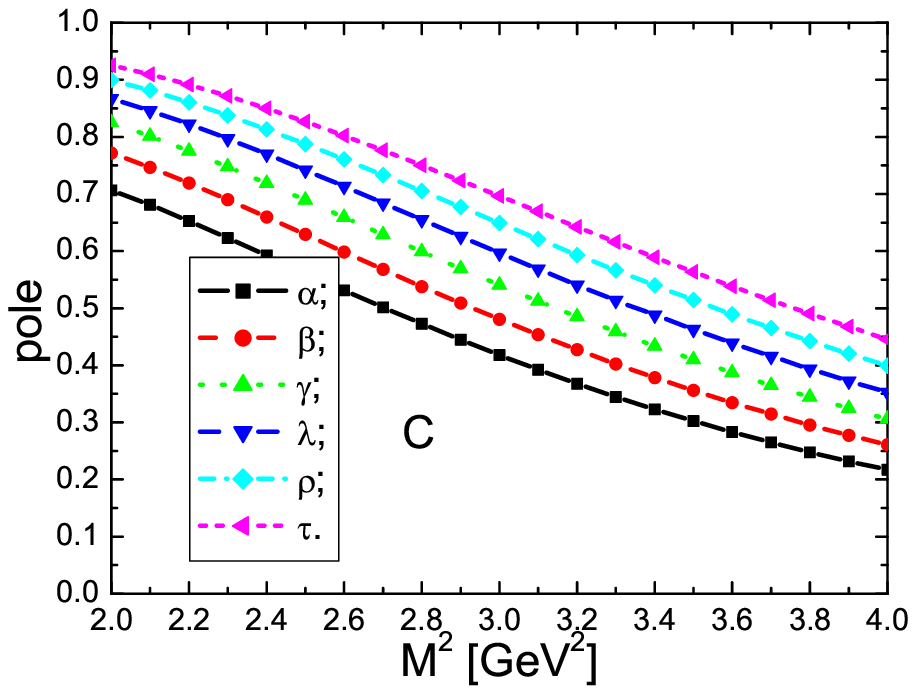}
 \includegraphics[totalheight=5cm,width=6cm]{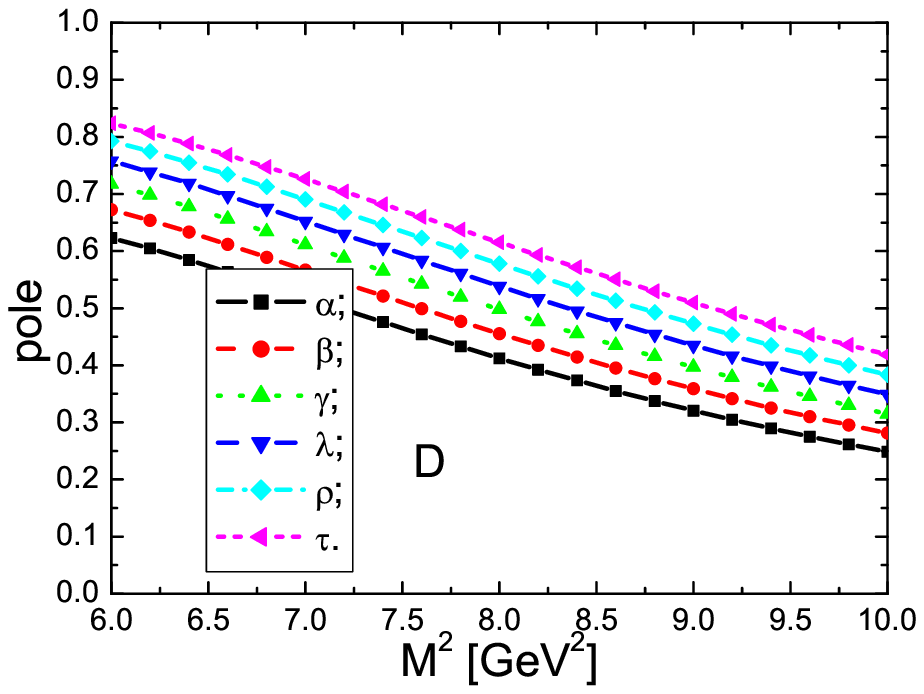}
 \includegraphics[totalheight=5cm,width=6cm]{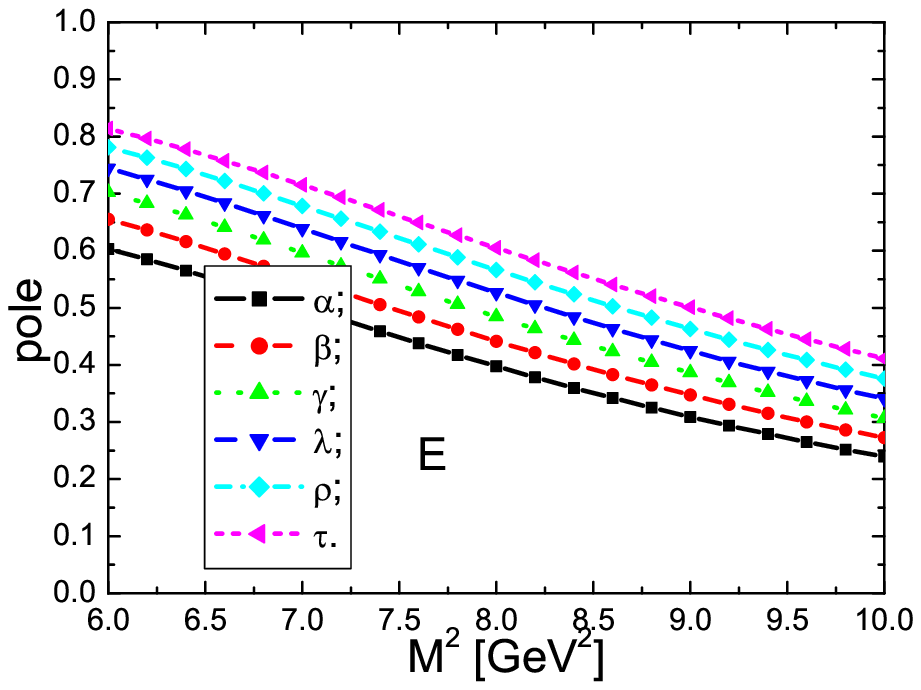}
 \includegraphics[totalheight=5cm,width=6cm]{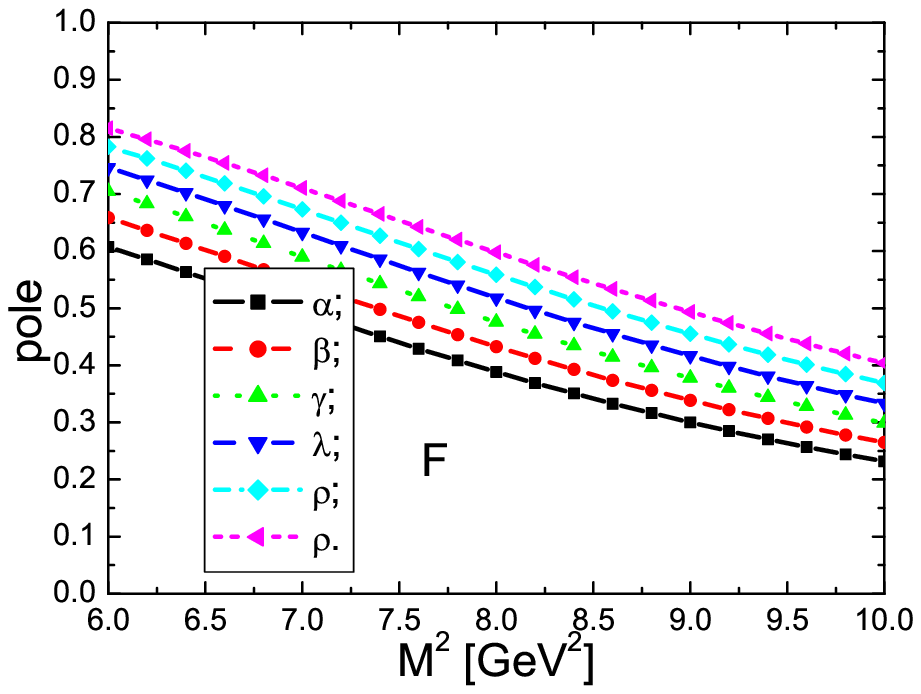}
   \caption{ The contributions from the pole terms with variation of the Borel parameter $M^2$. The $A$, $B$, $C$,
   $D$, $E$ and $F$ denote the $c\bar{c}q\bar{q}$,
   $c\bar{c}q\bar{s}$, $c\bar{c}s\bar{s}$, $b\bar{b}q\bar{q}$,
   $b\bar{b}q\bar{s}$ and $b\bar{b}s\bar{s}$ channels, respectively.  In the $c\bar{c}$ channels, the notations
   $\alpha$, $\beta$, $\gamma$, $\lambda$, $\rho$ and $\tau$  correspond to the threshold
   parameters $s_0=21\,\rm{GeV}^2$,
   $22\,\rm{GeV}^2$, $23\,\rm{GeV}^2$, $24\,\rm{GeV}^2$, $25\,\rm{GeV}^2$ and $26\,\rm{GeV}^2$ respectively
   ;  while in the $b\bar{b}$ channels they correspond to
    the threshold
   parameters  $s_0=132\,\rm{GeV}^2$,
   $134\,\rm{GeV}^2$, $136\,\rm{GeV}^2$, $138\,\rm{GeV}^2$, $140\,\rm{GeV}^2$ and $142\,\rm{GeV}^2$ respectively.}
\end{figure}

Taking into account all uncertainties of the input parameters,
finally we obtain the values of the masses and pole resides of
 the   $Z$, which are  shown in Figs.5-6 and Tables 1-2.

\begin{figure}
 \centering
 \includegraphics[totalheight=5cm,width=6cm]{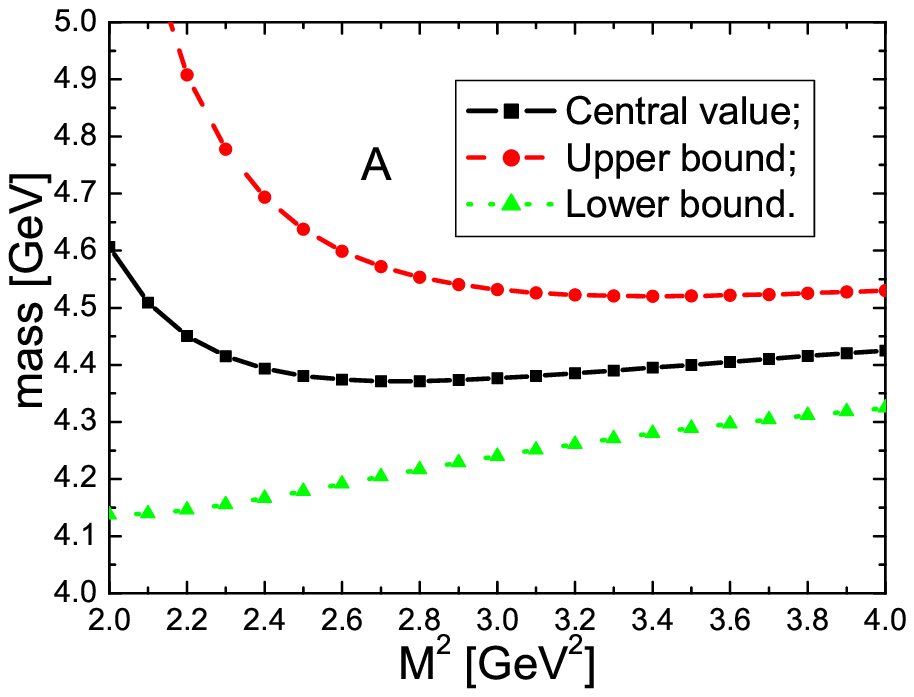}
 \includegraphics[totalheight=5cm,width=6cm]{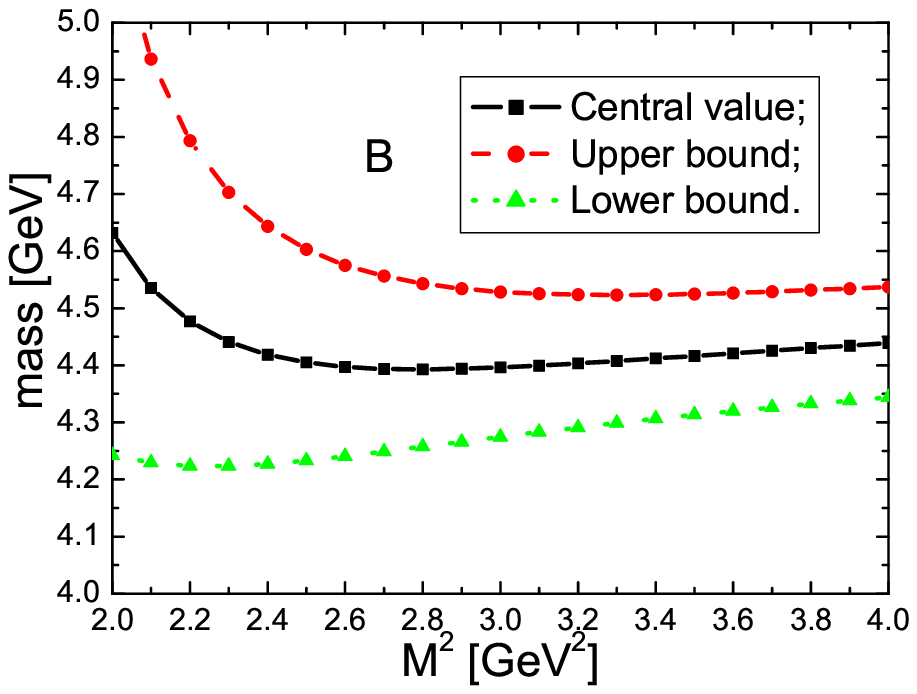}
 \includegraphics[totalheight=5cm,width=6cm]{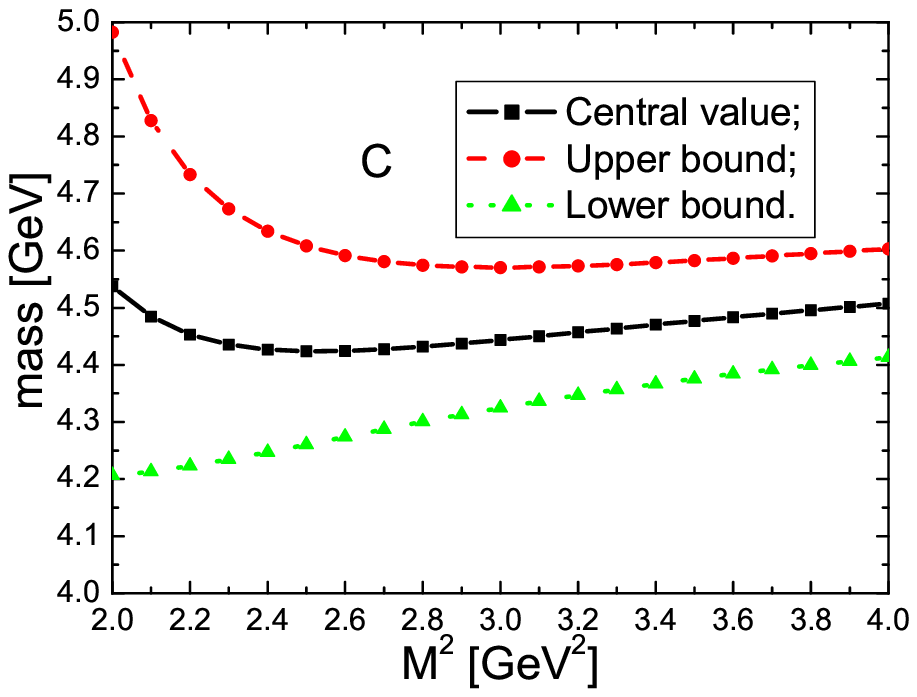}
 \includegraphics[totalheight=5cm,width=6cm]{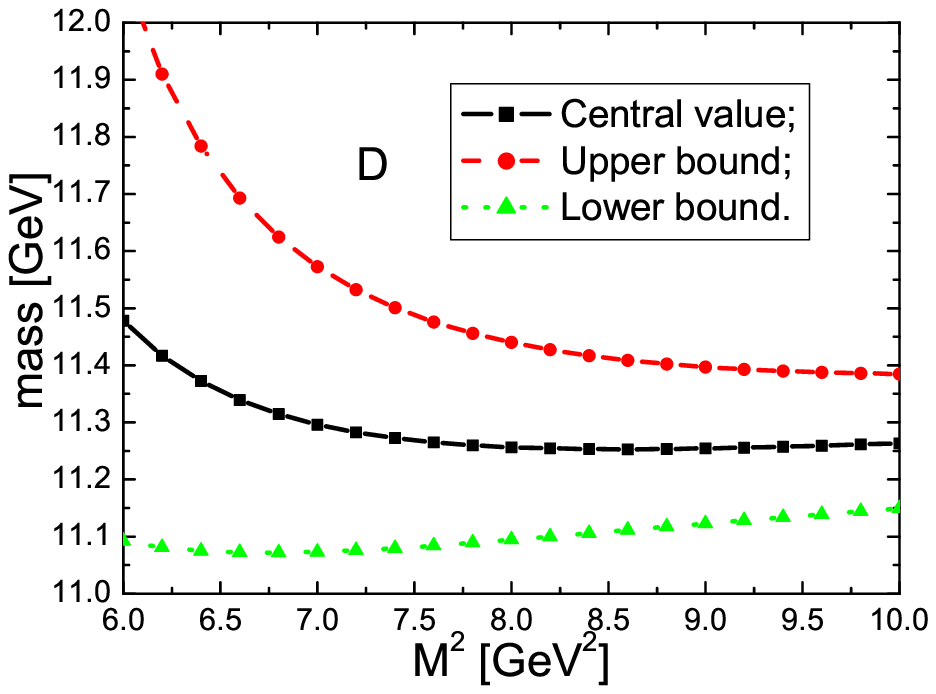}
 \includegraphics[totalheight=5cm,width=6cm]{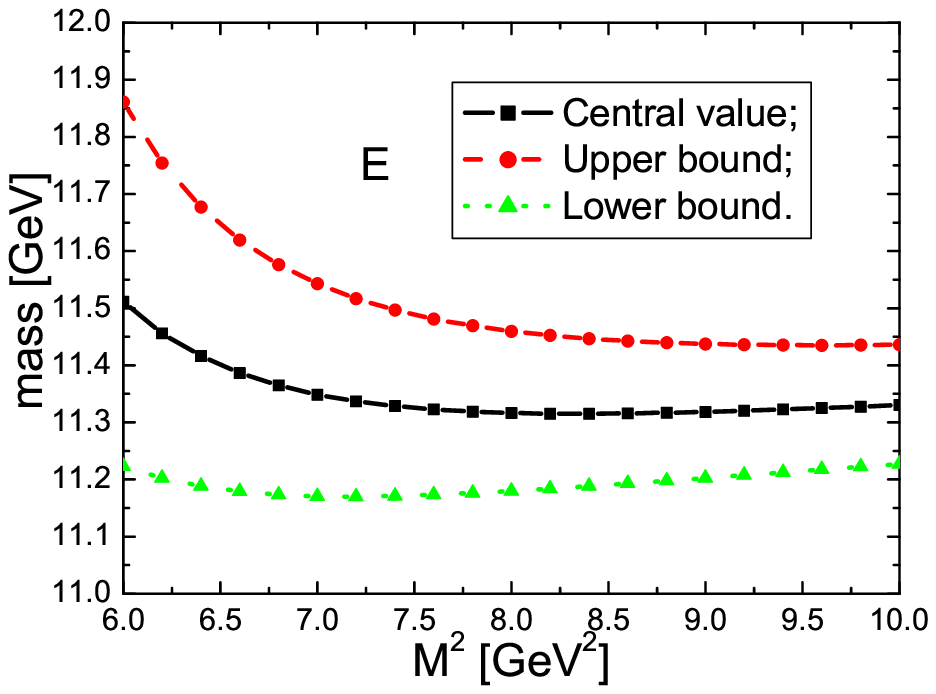}
 \includegraphics[totalheight=5cm,width=6cm]{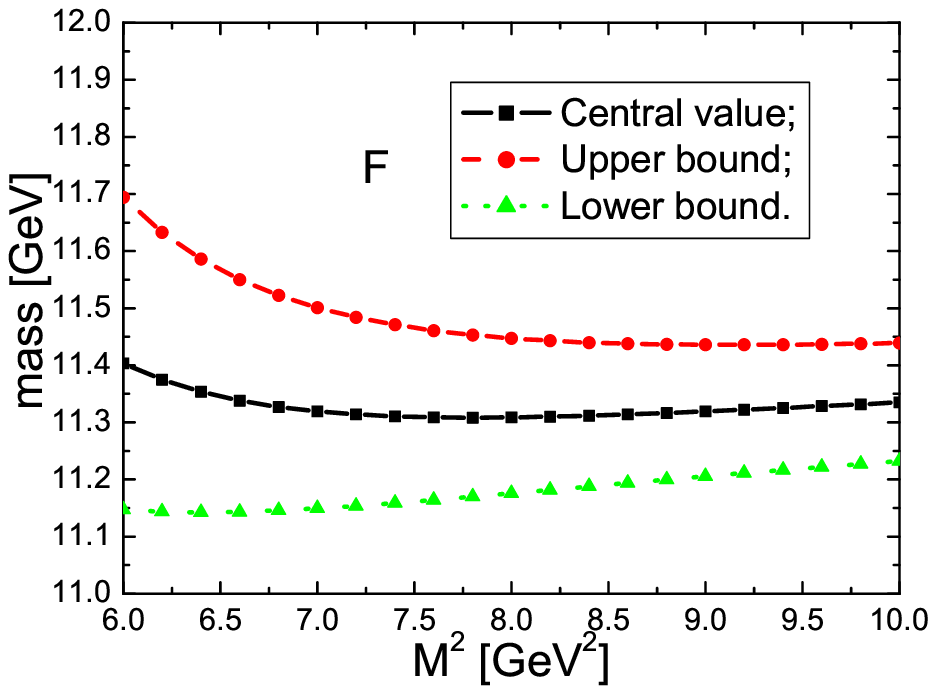}
   \caption{ The masses of the scalar tetraquark states with variation of the Borel parameter $M^2$. The $A$, $B$, $C$,
   $D$, $E$ and $F$ denote the $c\bar{c}q\bar{q}$,
   $c\bar{c}q\bar{s}$, $c\bar{c}s\bar{s}$, $b\bar{b}q\bar{q}$,
   $b\bar{b}q\bar{s}$ and $b\bar{b}s\bar{s}$ channels, respectively. }
\end{figure}

\begin{figure}
 \centering
 \includegraphics[totalheight=5cm,width=6cm]{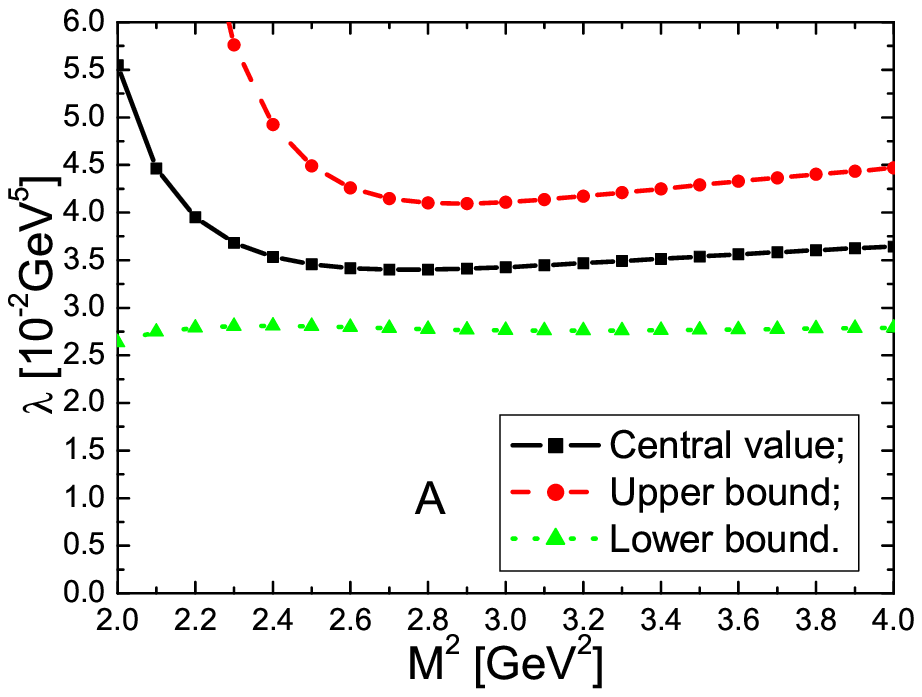}
 \includegraphics[totalheight=5cm,width=6cm]{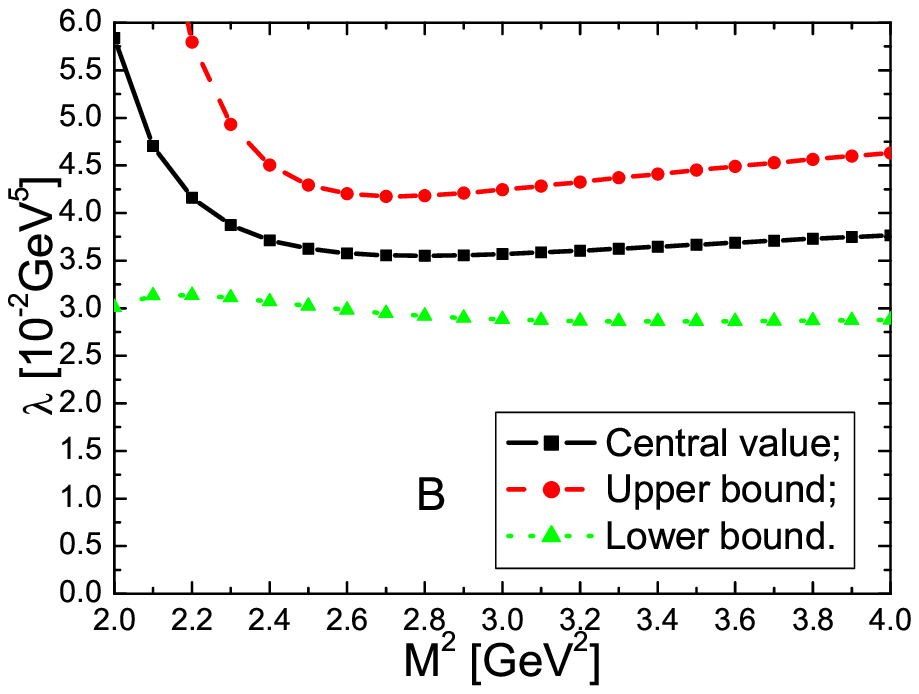}
 \includegraphics[totalheight=5cm,width=6cm]{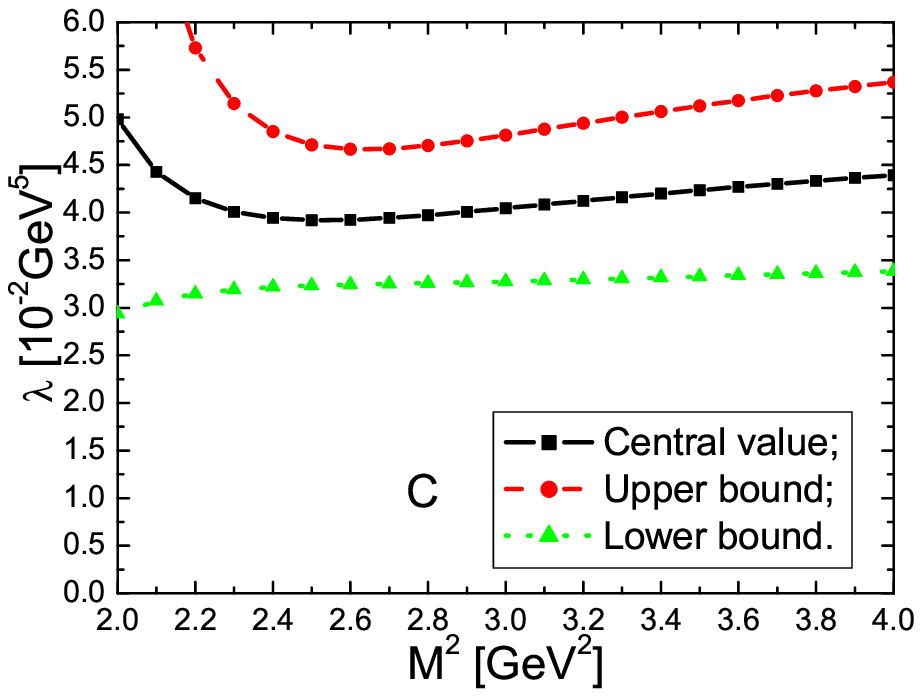}
 \includegraphics[totalheight=5cm,width=6cm]{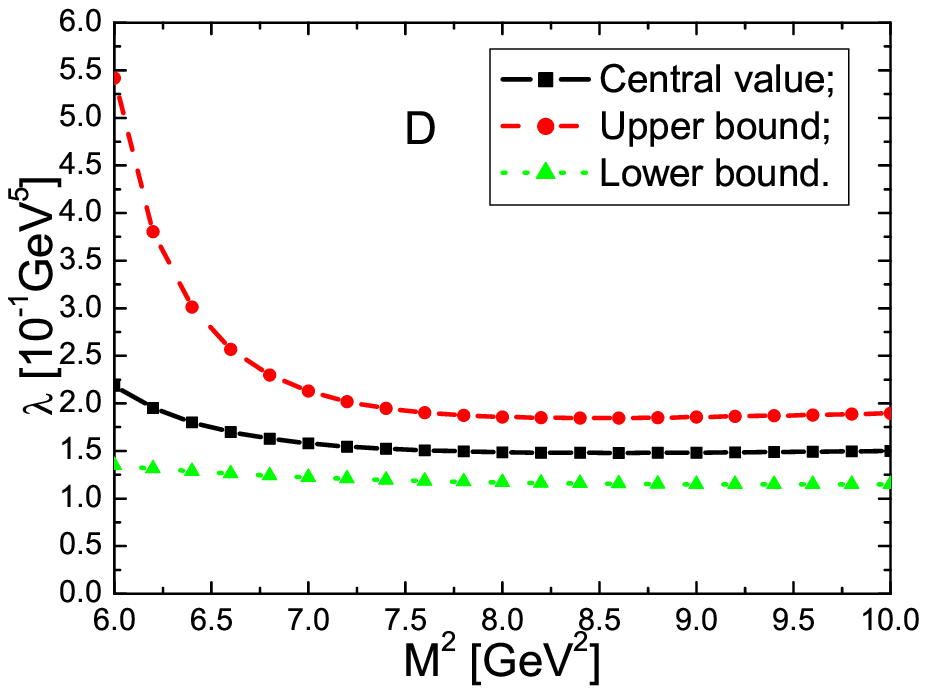}
 \includegraphics[totalheight=5cm,width=6cm]{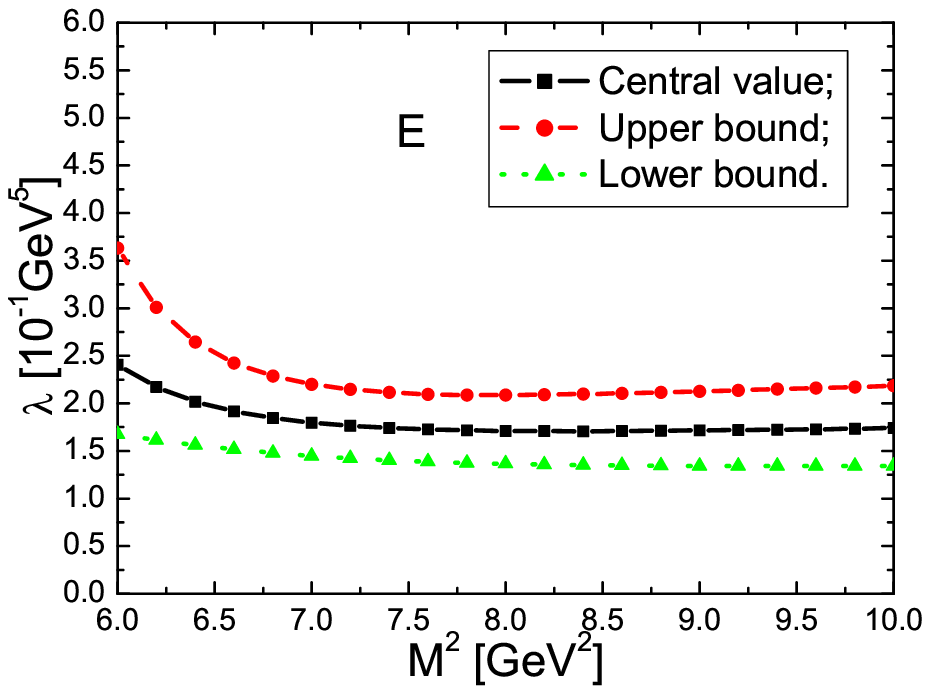}
 \includegraphics[totalheight=5cm,width=6cm]{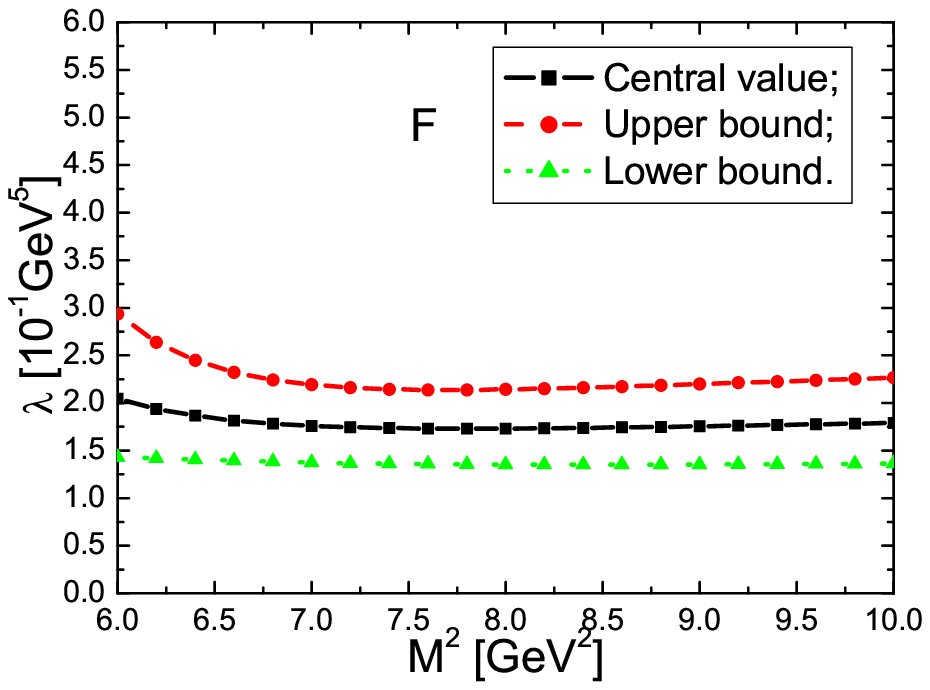}
   \caption{ The pole residues of the scalar tetraquark states with variation of the Borel parameter $M^2$. The $A$, $B$, $C$,
   $D$, $E$ and $F$ denote the $c\bar{c}q\bar{q}$,
   $c\bar{c}q\bar{s}$, $c\bar{c}s\bar{s}$, $b\bar{b}q\bar{q}$,
   $b\bar{b}q\bar{s}$ and $b\bar{b}s\bar{s}$ channels, respectively.}
\end{figure}

\begin{table}
\begin{center}
\begin{tabular}{|c|c|c|c|}
\hline\hline tetraquark states & This work & Refs.\cite{Ebert0512,Ebert0812}\\
\hline
      $c\bar{c}s\bar{s}$  &$4.44\pm0.16$ &$4.051$\\ \hline
       $ c\bar{c}q\bar{s}$& $4.39\pm0.16$& $3.922$\\     \hline
      $c\bar{c}q\bar{q} $ &$4.37\pm0.18$& $3.812$\\      \hline
    $b\bar{b}s\bar{s}$  &$11.31\pm0.16$ &$10.662$\\ \hline
       $ b\bar{b}q\bar{s}$& $11.33\pm0.16$ &$10.572$\\     \hline
      $ b\bar{b}q\bar{q} $ &$11.27\pm0.20$ &$10.471$\\      \hline
    \hline
\end{tabular}
\end{center}
\caption{ The masses (in unit of GeV)  for the scalar tetraquark
states. }
\end{table}

\begin{table}
\begin{center}
\begin{tabular}{|c|c|c|}
\hline\hline tetraquark states &pole residues ($10^{-2} \rm{GeV}^5$)\\
\hline
      $c\bar{c}s\bar{s}$  &$4.00\pm0.80$\\ \hline
       $ c\bar{c}q\bar{s}$ &$3.56\pm0.70$\\     \hline
      $c\bar{c}q\bar{q} $ & $3.41\pm0.70$\\      \hline
    $b\bar{b}s\bar{s}$   &$17.3\pm4.0$\\ \hline
       $ b\bar{b}q\bar{s}$ &$17.4\pm4.0$\\     \hline
      $ b\bar{b}q\bar{q} $  &$15.2\pm3.8$\\      \hline
    \hline
\end{tabular}
\end{center}
\caption{ The  pole residues for the scalar tetraquark states. }
\end{table}

From Table 1, we can see that the $SU(3)$ breaking effects for the
masses of the hidden charm and bottom tetraquark states are buried
in the uncertainties. The central value of the scalar tetraquark
state $c\bar{c}q\bar{q}$ is slightly larger than  the one
$M_{Z}=(4.36\pm0.18)\,\rm{GeV}$ obtained in Ref.\cite{Wang08072},
where the contributions from the terms involving the gluon
condensate are taken into account. We can draw the conclusion that
the gluon condensate plays a tiny important role and can be safely
neglected.

The  meson  $Z(4250)$   may be a scalar tetraquark state
($c\bar{c}u\bar{d}$), the decay $ Z(4250) \to \pi^+\chi_{c1}$ can
take place with the OZI super-allowed "fall-apart" mechanism, which
can take into account the large total width naturally.   Other
possibilities, such as a hadro-charmonium resonance and a
$D_1^+\bar{D}^0+ D^+\bar{D}_1^0$ molecular state are not excluded;
more experimental data are still needed to identify it. It is
difficult to identify  the $Z(4050)$    as the  scalar tetraquark
state ($c\bar{c}u\bar{d}$) considering its small mass. There still
lack experiential candidates to identify the scalar tetraquark
states $c\bar{c}q\bar{s}$, $c\bar{c}s\bar{s}$, $b\bar{b}q\bar{q}$,
$b\bar{b}q\bar{s}$ and $b\bar{b}s\bar{s}$.

In Table 1, we also present the results from a relativistic quark
model based on a quasipotential approach in QCD
\cite{Ebert0512,Ebert0812}, the central values of our predictions
are larger than the corresponding ones from the quasipotential model
about $(0.4-0.7)\, \rm{GeV}$. In Refs.\cite{Ebert0512,Ebert0812},
Ebert et al take the diquarks as  bound states of the light and
heavy quarks in the color antitriplet channel, and calculate their
mass spectrum using a Schrodinger type equation, then take the
masses of the diquarks  as the basic input parameters, and study the
mass spectrum of the heavy tetraquark states as bound states of the
diquark-antidiquark system. In the conventional quark models, the
constituent quark masses  are taken as the basic input parameters,
and fitted to reproduce the mass spectra  of the well known  mesons
and baryons. However, the present experimental knowledge about the
phenomenological hadronic spectral densities of the tetraquark
states is  rather vague, whether or not there exist   tetraquark
states is not confirmed with confidence, and no knowledge about  the
high resonances. The predicted constituent diquark masses can not be
confronted with the experimental data.

In Refs.\cite{Maiani20042,Maiani2008,Polosa0902}, Maiani et al take
the diquarks as the basic constituents, examine the rich spectrum of
the  diquark-antidiquark states  with  the constituent diquark
masses and the spin-spin
 interactions, and try to  accommodate some of the newly observed charmonium-like resonances not
 fitting a pure $c\bar{c}$ assignment. The predictions depend heavily on  the assumption that the light
 scalar mesons $a_0(980)$ and $f_0(980)$ are tetraquark states,
 the  basic  parameters (constituent diquark masses) are
 estimated thereafter. The predications  $M_{c\bar{c}q\bar{q}}=3723\,\rm{MeV}$
 \cite{Maiani20042} and $M_{c\bar{c}s\bar{s}}=3834\,\rm{MeV}$ \cite{Polosa0902} (for the
 tetraquark states $c\bar{c}q\bar{q}$   and $c\bar{c}s\bar{s}$
 respectively) are about $0.6\,\rm{GeV}$ smaller than the corresponding ones in the present work.

In Ref.\cite{Zouzou86}, Zouzou et al  solve the four-body ($\bar{Q}
\bar{Q}qq$) problem by three different variational methods with a
non-relativistic potential considering explicitly virtual
meson-meson components in the wave-functions, search for possible
bound states below the threshold for the spontaneous dissociation
into two mesons, and observe that the exotic bound states
$\bar{Q}\bar{Q}qq$ maybe exist for  unequal quark masses (the ratio
$m_Q/m_q$ is large enough). The studies  using a potential derived
from the MIT bag model in the Born-Oppenheimer approximation support
this observation \cite{Heller87,Heller88}. In Ref.\cite{Manohar93},
Manohar and Wise study systems of two heavy-light  mesons
interacting through  an one-pion exchange potential determined by
the heavy meson chiral perturbation theory and observe the long
range potential maybe sufficiently attractive to produce a weakly
bound two-meson state in the case $Q=b$. In
Ref.\cite{Silvestre-Brac}, the $L=0$ tetraquark states
$QQ\overline{QQ}$ ($Q$ denotes both $Q$ and $q$) are analyzed in  a
chromo-magnetic model where only a constant hyperfine potential   is
retained.

If there exist scalar tetraquark states $\bar{Q} \bar{Q}qq$, we can
construct the $C\gamma_\mu-C\gamma^\mu$ type interpolating currents
to study them with the QCD sum rules, as the $\bar{Q} \bar{Q}$ and
$qq$ favor forming  diquarks in the symmetric sextet $6_f$ with the
spin-parity  $J^P=1^+$ due to Fermi statistics. The attractive
interactions of one-gluon exchange  favor  formation of the diquarks
in  color antitriplet $\overline{3}_{ c}$, flavor antitriplet
$\overline{3}_{ f}$ and spin singlet $1_s$ \cite{GI1,GI2}. We expect
the scalar $C\gamma_\mu-C\gamma^\mu$ type tetraquark states $\bar{Q}
\bar{Q}qq$ are heavier  than the corresponding $C\gamma_5-C\gamma_5$
type tetraquark states $\bar{Q}Q \bar{q}q$, our numerical results
support this conjecture; our works on the $C\gamma_\mu-C\gamma^\mu$
type tetraquark states $\bar{Q} \bar{Q}qq$ will be presented
elsewhere.

The nonet scalar mesons below $1\,\rm{GeV}$ (the $f_0(980)$ and
$a_0(980)$ especially) are good candidates for the tetraquark
states. However, they can't satisfy the two criteria of the QCD sum
rules, and result in a reasonable Borel window.  If the perturbative
terms have the main contribution (in the conventional QCD sum rules,
the perturbative terms always have the main contribution), we can
approximate the spectral density with the perturbative term
\cite{Wang0708}, then take the pole dominance condition, and obtain
the approximate  relation,
\begin{eqnarray}
\frac{s_0}{M^2}\geq 4.7 \, .
\end{eqnarray}
If we take the Borel parameter has the typical value $M^2=1\,
\rm{GeV}^2$, then $s_0\geq 4.7\, \rm{GeV}^2$, the threshold
parameter is too large for the light tetraquark state candidates
$f_0(980)$, $a_0(980)$, etc.

On the other hand, the numerous candidates  with the same quantum
numbers $J^{PC}=0^{++}$ below $2 \,\rm{GeV}$ can't be accommodated
in one $q\bar{q}$ nonet,  some are supposed to be glueballs,
molecules and multiquark states
\cite{Jaffe2004,Close2002,ReviewScalar}. Once the main Fock sates of
the nonet scalar mesons below $1\,\rm{GeV}^2$ are proved  to be
tetraquark states, we can draw the conclusion that the QCD sum rules
are not applicable for the light tetraquark states.

In this article, we calculate the mass spectrum of the scalar
hidden charm and bottom tetraquark states  by imposing the two
criteria of the QCD sum rules. In fact, we usually  consult the
experimental data in choosing the Borel parameter $M^2$ and the
threshold parameter $s_0$. There lack experimental data for the
phenomenological hadronic spectral densities of the tetraquark
states, the present predictions can't be confronted with the
experimental data.

The LHCb is a dedicated $b$ and $c$-physics precision experiment at
the LHC (large hadron collider). The LHC will be the world's most
copious  source of the $b$ hadrons, and  a complete spectrum of the
$b$ hadrons will be available through gluon fusion. In proton-proton
collisions at $\sqrt{s}=14\,\rm{TeV}$¡Ì, the $b\bar{b}$ cross
section is expected to be $\sim 500\mu b$ producing $10^{12}$
$b\bar{b}$ pairs in a standard  year of running at the LHCb
operational luminosity of $2\times10^{32} \rm{cm}^{-2}
\rm{sec}^{-1}$ \cite{LHC}. The scalar tetraquark states predicted in
the present work may be observed at the LHCb, if they exist  indeed.
We can search for the scalar hidden charm tetraquark states  in the
$D\bar{D}$, $D^*\bar{D^*}$, $D_s\bar{D_s}$, $D_s^*\bar{D_s^*}$,
$J/\psi \rho$, $J/\psi \phi$, $J/\psi \omega$, $\eta_c\pi$,
$\eta_c\eta$, $\cdots$ invariant mass distributions and search for
the scalar hidden bottom tetraquark states  in the $B\bar{B}$,
$B^*\bar{B^*}$, $B_s\bar{B_s}$, $B_s^*\bar{B_s^*}$, $\Upsilon \rho$,
$\Upsilon \phi$, $\Upsilon \omega$, $\eta_b\pi$, $\eta_b\eta$,
$\cdots$ invariant mass distributions.

Furthermore, the non-leptonic $B$ decays through $ b\rightarrow c
\bar{c}s$  provide  another favorable environment for the production
of the scalar hidden charm tetraquark states \cite{Bigi2005}, we can
search for them at the KEK-B or the Fermi-lab Tevatron.

\section{Conclusion}
In this article, we study the mass spectrum of the scalar hidden
charm and bottom tetraquark states  with the QCD sum rules. The
numerical results are compared with the corresponding ones from a
relativistic quark model based on a quasipotential approach in QCD.
The relevant  values from the constituent diquark model
 based  on  the constituent diquark masses and the spin-spin
 interactions are also discussed. We can search for the scalar hidden
charm and bottom tetraquark states at the LHCb, the KEK-B or the
Fermi-lab Tevatron.

\section*{Acknowledgements}
This  work is supported by National Natural Science Foundation,
Grant Number 10775051, and Program for New Century Excellent Talents
in University, Grant Number NCET-07-0282.


\begin{thebibliography}{99}


\bibitem{review1} E. S. Swanson, Phys. Rept. {\bf 429} (2006) 243.

\bibitem{review2} E. Klempt and A. Zaitsev, Phys. Rept. {\bf 454} (2007) 1.

\bibitem{review3} M. B. Voloshin, Prog. Part. Nucl. Phys. {\bf 61} (2008) 455.


\bibitem{review4} S. Godfrey and  S. L. Olsen, Ann. Rev. Nucl. Part. Sci. {\bf 58} (2008) 51.

\bibitem{Olsen2009}  S. L. Olsen,  arXiv:0901.2371.

\bibitem{Belle-z4430} K. Abe et al, Phys. Rev. Lett. {\bf 100} (2008) 142001.


 \bibitem{Babar0811} B. Aubert et al, arXiv:0811.0564.

\bibitem{Belle-chipi}  R. Mizuk  et al,  Phys. Rev. {\bf D78} (2008) 072004.



\bibitem{Wang0807} Z. G. Wang, Eur. Phys. J. {\bf C59} (2009) 675.

\bibitem{Wang08072} Z. G. Wang, arXiv:0807.4592.

\bibitem{Xliu0808} X. Liu, Z. G. Luo, Y. R. Liu  and S. L. Zhu,
arXiv:0808.0073.

\bibitem{Lee09} S. H. Lee, K. Morita and M. Nielsen, Nucl. Phys. {\bf A815} (2009) 29.

\bibitem{SLee08} S. H. Lee, K. Morita and M. Nielsen, Phys. Rev. {\bf D78} (2008) 076001.

\bibitem{GDing09} G. J. Ding, Phys. Rev. {\bf D79} (2009) 014001.


\bibitem{SVZ79}  M. A. Shifman, A. I. Vainshtein and V. I. Zakharov,
Nucl. Phys. {\bf B147} (1979) 385, 448.

\bibitem{Reinders85} L. J. Reinders, H. Rubinstein and S. Yazaki, Phys. Rept. {\bf 127} (1985) 1.

\bibitem{Jaffe2003} R. L. Jaffe and  F. Wilczek, Phys. Rev. Lett. {\bf 91} (2003) 232003.


\bibitem{Jaffe2004} R. L. Jaffe, Phys. Rept. {\bf 409} (2005) 1.

\bibitem{GI1} A. De Rujula, H. Georgi and S. L. Glashow, Phys. Rev.  {\bf D12}
(1975) 147.

\bibitem{GI2} T. DeGrand, R. L. Jaffe, K. Johnson and J. E. Kiskis,
Phys.  Rev.  {\bf D12} (1975) 2060.


\bibitem{Wang1} Z. G. Wang, Nucl. Phys. {\bf A791} (2007) 106.


\bibitem{Wang2} Z. G. Wang, W. M. Yang and S. L. Wan, J. Phys. {\bf G31} (2005) 971.

\bibitem{Wang0904} Z. G. Wang, arXiv:0903.5200.

\bibitem{Ioffe2005} B. L. Ioffe, Prog. Part. Nucl. Phys. {\bf 56} (2006)
232.

\bibitem{Wang0708} Z. G. Wang, Chin. Phys. {\bf C32} (2008) 797.

\bibitem{Ebert0512} D. Ebert, R. N. Faustov and V. O. Galkin, Phys. Lett. {\bf B634} (2006) 214.


\bibitem{Ebert0812}  D. Ebert, R. N. Faustov and V. O. Galkin, arXiv:0812.3477.



\bibitem{Maiani20042} L. Maiani, F. Piccinini, A. D. Polosa and V. Riquer, Phys. Rev. {\bf D71} (2005) 014028.

\bibitem{Maiani2008} L. Maiani, A. D. Polosa and V. Riquer, New J. Phys. {\bf 10} (2008) 073004.

\bibitem{Polosa0902}  N. V. Drenska, R. Faccini and  A. D. Polosa, arXiv:0902.2803.

\bibitem{Zouzou86} S. Zouzou, B. Silvestre-Brac, C. Gignoux and J. M. Richard, Z. Phys. {\bf C30} (1986) 457.

\bibitem{Heller87} L. Heller and J. A. Tjon, Phys. Rev. {\bf D35} (1987) 969.

\bibitem{Heller88} J. Carlson, L. Heller and J. A. Tjon, Phys. Rev. {\bf D37} (1988)
744.

\bibitem{Manohar93}  A. V. Manohar and M. B. Wise, Nucl. Phys. {\bf  B399} (1993)
17.

\bibitem{Silvestre-Brac} B. Silvestre-Brac, Phys. Rev. {\bf D46} (1992)  2179.

\bibitem{Close2002} F. E. Close and N. A. Tornqvist, J. Phys.  {\bf G28} (2002) R249.

\bibitem{ReviewScalar}   C. Amsler and N. A. Tornqvist, Phys. Rept. {\bf 389} (2004) 61.


\bibitem{LHC}  G. Kane and A. Pierce, "Perspectives On LHC Physics",
World Scientific Publishing Company,  2008.

\bibitem{Bigi2005} I. Bigi, L. Maiani, F. Piccinini, A. D. Polosa and V. Riquer, Phys. Rev. {\bf D72} (2005) 114016.




\end{thebibliography}
\end{document}